\documentclass[%
 reprint,
preprintnumbers,
 amsmath,amssymb,
 aps,
 prc,
]{revtex4-2}

\usepackage{graphicx} % Include figure files
\usepackage{dcolumn}  % Align table columns on decimal point
\usepackage{bm}       % Bold math
\usepackage{xcolor}   % Colors
\usepackage{color}

\graphicspath{{./}{figs/}}
\usepackage{graphicx}
\usepackage[tight,footnotesize]{subfigure}

\begin{document}

\preprint{LA-UR-22-30081}

\title{Collective enhancement in the exciton model}% Force line breaks with \\
%\thanks{A footnote to the article title}%

% Alphabetical ordering for now

\author{M. R. Mumpower}
\email{mumpower@lanl.gov}
\affiliation{Los Alamos National Laboratory, Los Alamos, NM, 87545, USA}

\author{D. Neudecker}
%\email{Author@institution.gov}
\affiliation{Los Alamos National Laboratory, Los Alamos, NM, 87545, USA}

\author{H. Sasaki}
%\email{Author@institution.gov}
\affiliation{Los Alamos National Laboratory, Los Alamos, NM, 87545, USA}

\author{T. Kawano}
%\email{Author@institution.gov}
\affiliation{Los Alamos National Laboratory, Los Alamos, NM, 87545, USA}

\author{A. E. Lovell}
%\email{Author@institution.gov}
\affiliation{Los Alamos National Laboratory, Los Alamos, NM, 87545, USA}

\author{M. W. Herman}
%\email{Author@institution.gov}
\affiliation{Los Alamos National Laboratory, Los Alamos, NM, 87545, USA}

\author{I. Stetcu}
%\email{Author@institution.gov}
\affiliation{Los Alamos National Laboratory, Los Alamos, NM, 87545, USA}

\author{M. Dupuis}
%\email{Author@institution.gov}
\affiliation{CEA, DAM, DIF, F-91297 Arpajon, France}
\affiliation{Universit\'{e} Paris-Saclay, CEA, Laboratoire Mati\`{e}re sous Conditions Extr\^{e}mes, 91680 Bruy\`{e}res-Le-Ch\^{a}tel, France}

%\author{Co Authors...}%
%\affiliation{Institute}

\date{\today}% It is always \today, today,
             %  but any date may be explicitly specified

\begin{abstract}
The pre-equilibrium reaction mechanism is considered in the context of the exciton model. 
A modification to the one-particle one-hole state density is studied which can be interpreted as a collective enhancement. 
The magnitude of the collective enhancement is set by simulating the Lawrence Livermore National Laboratory (LLNL) pulsed-spheres neutron-leakage spectra. 
The impact of the collective enhancement is explored in the context of the highly deformed actinide, $^{239}$Pu. 
A consequence of this enhancement is the removal of fictitious levels in the Distorted-Wave Born Approximation often used in modern nuclear reaction codes. 
\end{abstract}

%\keywords{Suggested keywords}%Use showkeys class option if keyword
                              %display desired
\maketitle

%\tableofcontents

% ========================
\section{Introduction} \label{sec:intro}

Nuclear reaction modeling for strongly deformed nuclei remains an open challenge for contemporary theoretical studies.
Modern reaction codes separate the reaction mechanisms into three broad categories. 
In a direct reaction, the incident particle interacts on a fast time scale with a single nucleon that generally resides near the surface of the target system. 
The direct reaction cross section evolves slowly as a function of incident particle energy \cite{Glendenning2004}. 
In contrast, compound nucleus formation occurs when a large number of nucleons participate for a sufficiently long enough time that a thermal equilibrium ensues in the residual system \cite{Loveland1972}. 
This mechanism occurs at low energies inside the volume of the residual system. 
The cross section of this mechanism may vary strongly with small change in the incident-particle energy. 

Pre-equilibrium is the third, intermediate reaction mechanism that embodies both direct- and compound-like features. 
Pre-equilibrium reactions occur on a longer timescale than a direct reaction but on a shorter timescale than compound nucleus formation \cite{Griffin1966}. 
This mechanism is characterized by an incident particle that continually enables subsequent scattering. 
As the scattering proceeds, increasingly more complex states are created in the residual system with each successive process gradually losing information contained in the initial reaction. 
This reaction mechanism is important to consider with highly energetic incident particles. 
If the residual system has sufficient excitation energy, creation of subsequent particles may be possible \cite{Gruppelaar1986}. 

There are two distinct approaches to describe the pre-equilibrium process for nucleon-induced reactions on medium- to heavy- mass nuclei: purely quantum mechanical models and phenomenological-based models. 
Quantum mechanical models use the Distorted-Wave Born Approximation (DWBA) for the multi-step process to couple to the continuum in a residual nucleus. 
These models adopt different statistical assumptions, mainly for the two-step process, where 2-particle-2-hole configurations are created by the $NN$ interaction. 
Examples of quantum mechanical models are Feshbach-Kerman-Koonin
(FKK)~\cite{Feshbach1980}, Tamura-Udagawa-Lenske (TUL)~\cite{Tamura1982},  Nishioka-Weidenm\"{u}ller-Yoshida (NWY)~\cite{Nishioka1989}, and Luo-Kawai~\cite{Luo19991}. 

Because the angular momentum conservation is properly included in these quantum mechanical models, they better reproduce the $\gamma$-ray production data that are sensitive to the spins of initial and final states~\cite{Kerveno2021}. 
While these models provide more fundamental insight into nuclear reaction mechanisms, the downside of their application in nucleon-induced reactions is their high computational cost for the description of the relatively small fraction of the total reaction cross section. 

The second approach is phenomenological in nature. 
An example is the exciton model which treats pre-equilibrium scattering as a chain of particle-hole excitations \cite{Ribansky1973, Gudima1983}. 
In this context, the particle and hole degrees of freedom are referred to collectively as excitons and the exciton number for a single component system is given by $n = p + h$. 
Transitions between particle-hole configurations with the same exciton number, $n$, have equal probability. 
The time-dependent master equations controls the evolution of the scattering process through transitions to more or less complex configurations. 
At any step in this process an outgoing particle may be emitted which is referred to as pre-equilibrium emission. 
The time integrated solution provides the energy averaged particle spectra. 
Central to the exciton model is the set of particle-hole state densities that govern the magnitude of the excitations. 
In particular, the relative magnitude of the state densities are not fully constrained by differential data. 

A practical step forward is to combine both of these approaches: feed the quantum mechanical calculations to the exciton model. 
For example, the angular momentum transferred to a 1-particle-1-hole configuration is calculated by FKK, and the spin distribution of the populated final states are parameterized in the exciton model~\cite{Kawano2006b,Dashdorj2007,Kerveno2021}. 
This technique enables a more realistic spin transfer to the residual nucleus, while the whole pre-equilibrium strength can be determined by the more established exciton model framework. 

Although this combined approach compensates deficient information of angular momentum transfer in the exciton model, it is insufficient to provide individual contributions from different particle-hole configurations to the total pre-equilibrium energy spectrum. 
It is known that deformed nuclei at relatively low excitation energies show collective behavior, which can be evaluated by the Quasi-particle Random Phase Approximation (QRPA)~\cite{Dupuis2015,Dupuis2017}, as shown by Kerveno \textit{et al.}~\cite{Kerveno2021}. 
This collective excitation can be interpreted as an effective enhancement in the partial state density for 1-particle-1-hole configurations. 
Ergo, incorporating a collective enhancement for the 1-particle-1-hole state density into the exciton model may offer better modeling of the entire nuclear reaction occurring in highly deformed nuclei such as the actinides. 
Crucially, this procedure can be integrated into the Hauser-Feshbach theory which follows the statistical decay of the residual nucleus. 

In this paper we study this combined practical technique. 
We propose an increase to the 1-particle-1-hole state density used in the exciton model and include it in the Los Alamos statistical model framework, CoH$_3$ \cite{Kawano2016, Kawano2019}. 
We study the impact of this enhancement in the context of neutron-induced reactions on $^{239}$Pu. 
We use feedback from Lawrence Livermore National Laboratory (LLNL) pulsed-sphere neutron-leakage spectra to set the magnitude of the enhancement factor and find that this scale factor is significantly above unity. 
We present the changes to the cross sections in the results section and summarize our findings in the final section. 

% ========================
\section{Theory}

% ------------------------
\subsection{Exciton model}

\begin{figure}
    \centering
    \includegraphics[width=0.45\textwidth]{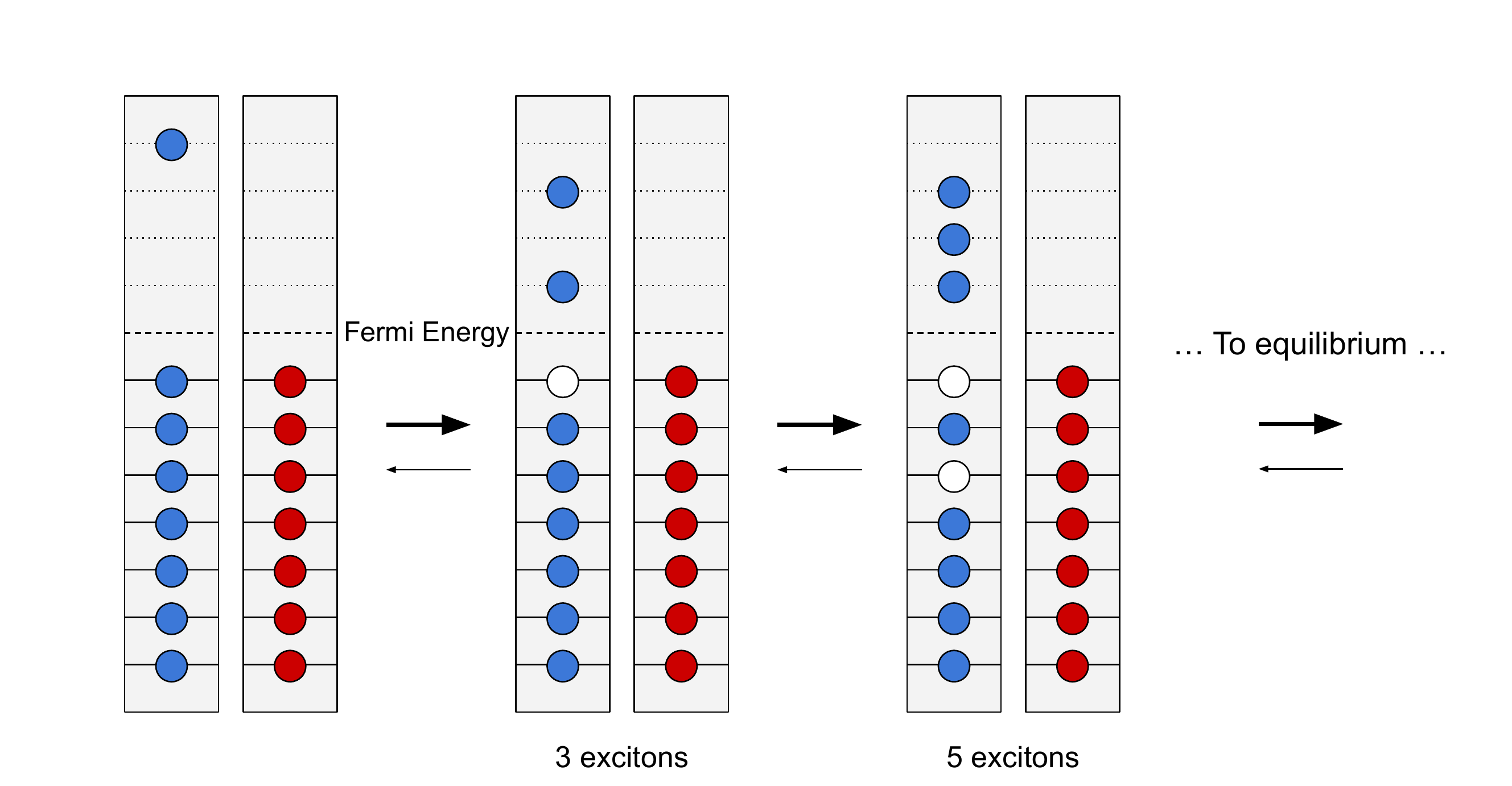}
    \caption{(Color Online) A schematic depiction of the first few stages of the 2-component exciton model from an initial excitation with a neutron. The particles, in this case nucleons (neutrons and protons), are shown as filled circles with holes indicated by open circles. The solid lines represent equally spaced single-particle states. }
    \label{fig:exiton_schematic}
\end{figure}

We employ the two-component exciton model~\cite{Kalbach1986, Koning2004}, which distinguishes neutron and proton in the particle-hole configurations. 
This is denoted schematically in Fig.~\ref{fig:exiton_schematic}. 
Since this model has been well established and extensively applied to particle emission data analysis, only a brief description of some of the relevant parts of the model is given below. 

We denote the particle-hole configuration by $c$, which abbreviates the number of particles and holes in the neutron and proton shells as $c \equiv (p_\nu, h_\nu, p_\pi, h_\pi)$. 
We also define the total number of excitons, $n_\nu = p_\nu + h_\nu$, $n_\pi = p_\pi + h_\pi$, and $n_t = n_\nu+n_\pi$. 
For a particle having $z$-protons and $n$-neutrons emitted in output channel $b$, the residual configuration will be designated by $c_b$, that stands for $p_\pi - z$ and $p_\nu-n$. 
In the case of an incident neutron on a target system with $Z$-protons and $N$-neutrons, the composite system would be the nucleus --- before compound nucleus formation --- ($Z$,$N+1$), and the residual system might be ($Z$, $N$) after emission of the neutron, e.g. in the case of inelastic scattering. 

For the pre-equilibrium nuclear reaction, $(a,b)$, with input channel, $a$, and output channel, $b$, the emission rate of the outgoing particle $b$ is written as
\begin{equation}
 W_b(c,E,\epsilon_b)
  = \frac{2s_b+1}{\pi^2 \hbar^3} \mu_b
    \sigma^{\rm CN}_b(\epsilon_b) \epsilon_b
    \frac{\omega(c_b,U)}{\omega(c,E)} f_{\rm FW} \ ,
 \label{eq:emissionrate}
\end{equation}
where $E$ is the total energy of the composite system, $U$ is the excitation energy in the residual nucleus, and $\omega(c,E)$ is the composite state density at the excitation energy $E$. 
A commonly used step function, $f_{\rm FW}$, is employed to limit the hole state configuration within the potential depth~\cite{Betak1976}. 
The values of $\epsilon$, $s_b$, and $\mu_b$, denote the emission energy, the intrinsic spin of particle $b$, and the reduced mass respectively. 
The compound formation cross section for the inverse reaction calculated by the particle
transmission coefficient is $\sigma^{\rm CN}_b(\epsilon_b)$. 

The pre-equilibrium emission takes place at different particle-hole configurations, which is characterized by the occupation probability $P(c)$ and its lifetime $\tau(c)$. 
The observed energy-differential cross section is a convolution of all the configurations
\begin{equation}
\frac{d\sigma}{d\epsilon_b} =
  \sigma^{\rm CN}_a(\epsilon_a)
  \sum_c P(c) \tau(c) W_b(c,E,\epsilon_b) \ ,
\end{equation}
where $\sigma^{\rm CN}_a$ is the compound nucleus formation cross section for channel $a$. 

We employ the $\tau(c)$ calculation proposed by Kalbach~\cite{Kalbach2006} and adopt the closed-form expression for $P(c)$. 
The most important ingredients of this model are the single-particle state densities, $g$, and the effective average squared matrix element $M^2$ for the two-body interaction. 
The effective average squared matrix element is considered as an adjustable model parameter in the exciton model, and often phenomenologically parameterized by comparing with experimental
data~\cite{Koning2004}. 
We now discuss the single-particle state densities (the $g$'s) and their role in setting the composite state density, $\omega$.

% ------------------------
\subsection{State density}

The composite state density $\omega(c,E)$ is given by the Williams' formula~\cite{Williams1971} for the two-component case~\cite{Kalbach1986}
\begin{equation}
 \omega(c,E)
  = \frac{g_\nu^{n_\nu} g_\pi^{n_\pi} \left \{ E - \Delta - A(c,E) \right\}^{n_t-1} }
         {p_\pi ! h_\pi ! p_\nu ! h_\nu ! (n_t - 1)!} \ ,
 \label{eq:omega}
\end{equation}
where $g_{\nu,\pi}$ is the single-particle state densities, $\Delta$ is
the pairing correction energy~\cite{ORNL-7042}, and $A(c,E)$ is the
Pauli correction function defined as,  
\begin{eqnarray}
  A(c,E) &=& E_{\rm th}
          - \frac{p_\nu^2 + h_\nu^2 + n_\nu}{4 g_\nu} 
          - \frac{p_\pi^2 + h_\pi^2 + n_\pi}{4 g_\pi} \ ,\\
  E_{\rm th} &=&
            \frac{[\max(p_\nu,h_\nu)]^2}{g_\nu} 
          + \frac{[\max(p_\pi,h_\pi)]^2}{g_\pi} \ .
\end{eqnarray}
This formula can be derived under the assumption that the single-particle states are equally spaced in energy. 
The single-particle state densities for the neutron and proton shell in Eq.~(\ref{eq:omega}) are often estimated simply by $g_\nu = N/C_\nu$ and $g_\pi = Z/C_\pi$, where $C_{\nu,\pi}$ is between 10 and 20~MeV. 
Values in the lower end of this range correspond to single particle levels are evenly distributed near the Fermi surface. 

In a more microscopic view, $g_{\nu}$ or $g_{\pi}$ can be evaluated by solving the Schr\"{o}dinger equation for a one-body potential, and applying Strutinsky's method~\cite{Strutinsky1968, Bolsterli1972} to extract the single particle state density. Alternative approach was proposed by Shlomo~\cite{Shlomo1991}.
Using the Strutinsky approach, we employ the axially-symmetric folding Yukawa potential of the finite range droplet model (FRDM)~\cite{Moller1995, Moller2016} to generate the single-particle state density $g(\varepsilon)$ for various nuclei:
\begin{equation}
  g(\varepsilon) = \sum_{i} \delta(\varepsilon - \varepsilon_i) , 
\end{equation}
where $\varepsilon_i$ is the energy of $i$-th single-particle state in the folding Yukawa potential. 
The single-particle state density, $g(\varepsilon)$, is expanded by a series of the Hermite polynomial to separate into a smoothly varying part $\overline{g}(\varepsilon)$ and locally fluctuating part $\delta
g(\varepsilon)$~\cite{Bolsterli1972}. 

\begin{figure}
  \includegraphics[width=0.45\textwidth]{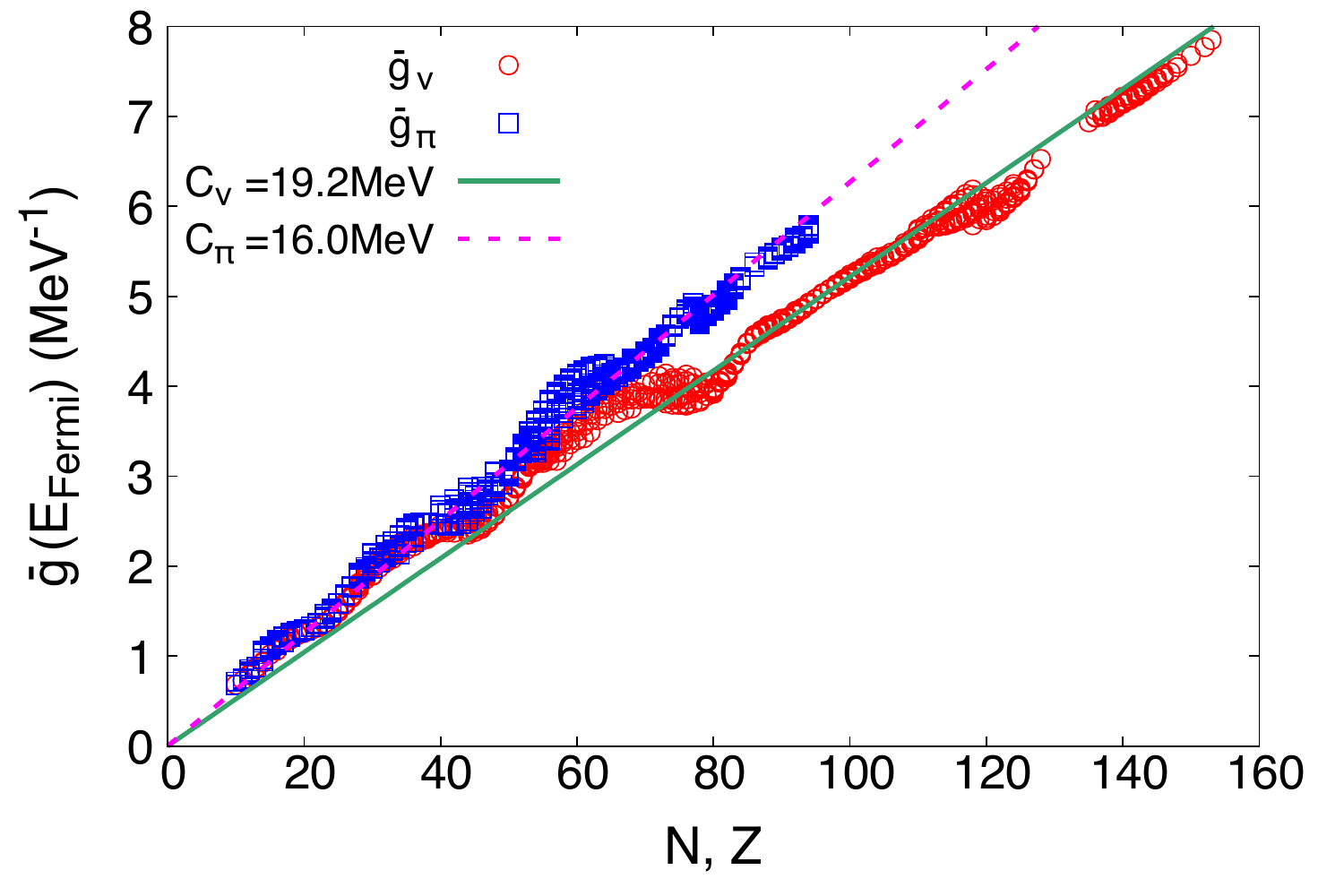}
  \caption{Single-particle state densities for various nuclei calculated for the axially-symmetric folding Yukawa potential. The solid and dashed lines are fitted lines.}
  \label{fig:spdensity}
\end{figure}

The single-particle state density at the Fermi surface is given by $\overline{g}(\varepsilon=E_{\rm Fermi})$. 
This quantity calculated from FRDM is shown for a range of stable nuclei in Fig.~\ref{fig:spdensity}. 
Also shown are the linear approximations to $\bar{g}_{\nu}$ and $\bar{g}_{\pi}$ using $C_\nu=19.2$ and $C_\pi=16.0$~MeV. 
The single-particle state density for neutrons is generally found to be less than that of protons for the same number of particles as indicative of the aforementioned constants. 

The state density for a system can also be constructed via a combinatorial method for the single-particle, ${\varepsilon_i}$ spectrum. 
For completeness, we also perform this calculation, following the work of Ref.~\cite{Herman1987} and references therein, in which all the $1p$-$1h$ configurations are combined using the single-particle levels from FRDM. 
Unlike the assumptions of the Williams' formula [Eq.~(\ref{eq:omega})], these single-particle states are not equally spaced in energy. 
As an approximation to the residual system, $^{239}$Pu, we use the composite system, $^{240}$Pu to compute the $1p$-$1h$ state density, where axial symmetric deformation is assumed without a unpaired nucleon. The difference in the particle-hole level densities in these nuclei is negligible.

\begin{figure}
  \begin{center}
    \resizebox{\columnwidth}{!}{\includegraphics{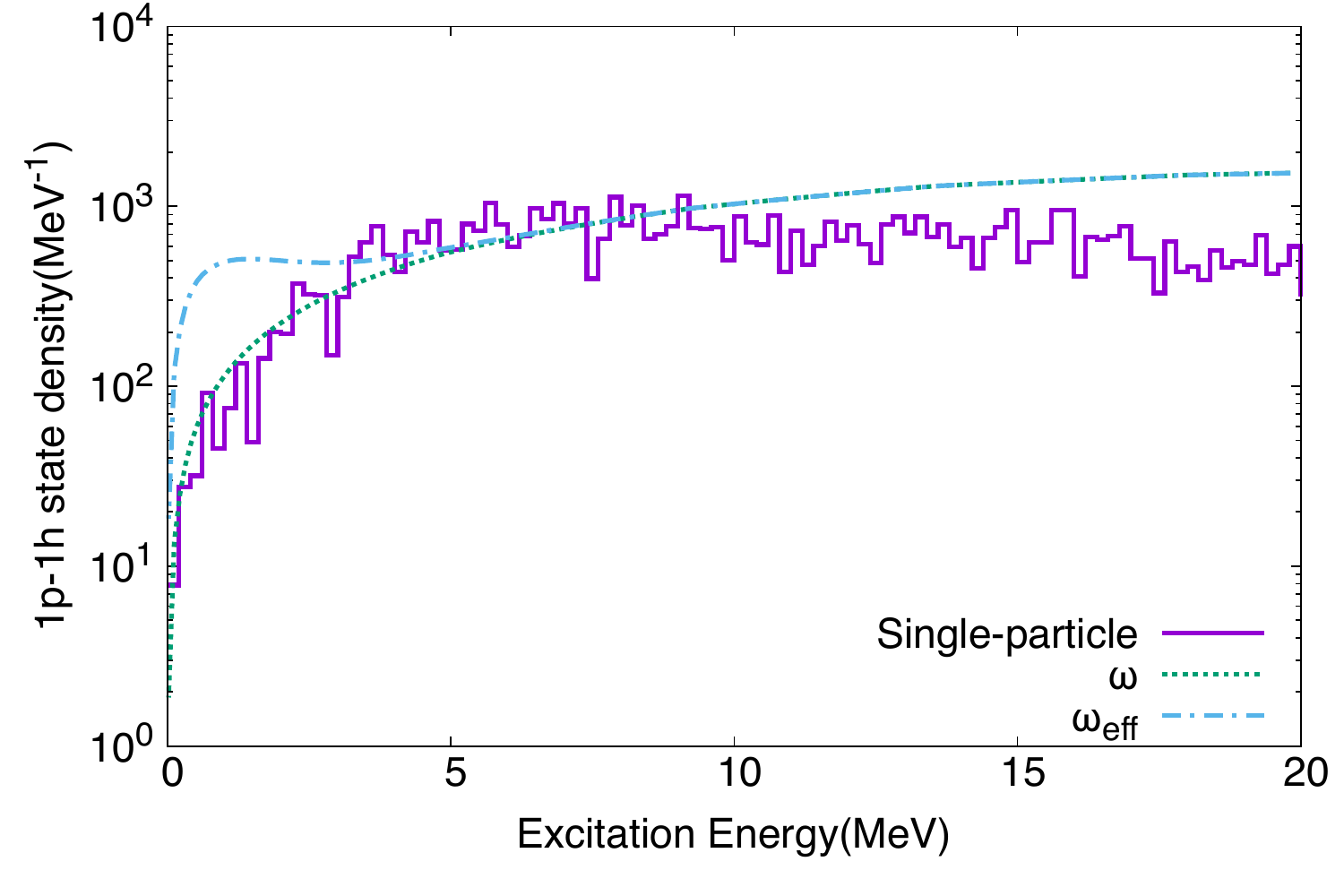}}
  \end{center}
  \caption{
  The $1p$-$1h$ state densities for $^{239}$Pu calculated by Williams' formula~\cite{Williams1971} (dotted and dot-dashed lines) and $^{240}$Pu calculated by the combinatorial single-particle model (solid line). The difference between $^{239}$Pu and $^{240}$Pu is negligible.}
  \label{fig:phdensity}
\end{figure}

Figure ~\ref{fig:phdensity} shows the $1p$-$1h$ state density calculated combinatorially from the single-particle states (solid line) and from application of the Williams' formula (dotted line). 
Good agreement is found between these two approaches, especially between an excitation energy of 0 to 10 MeV. 
Above 10 MeV, both calculations flatten out where other higher order $p$-$h$ excitations dominate. 

% ------------------------
\subsection{Collective enhancement of state density}
The emission rate $W_b(c,E,\epsilon_b)$ is proportional to $\omega(c_b,U)dU$, which is the number of final states in the residual nucleus. 
In the case of neutron inelastic scattering, the composite system the $p$-$h$ configuration can be either $c = (2,1,0,0)$ or $c = (1,0,1,1)$, while the residual system $c_b = (1,1,0,0)$ or $(0,0,1,1)$.  
The $1p$-$1h$ state density in Eq.~(\ref{eq:omega}) of the residual system reduces to $(g_\nu^2 g_\pi^2)(E-\Delta)$ at relatively low excitation energy and it is in this region that the change to the state density will be explored. 

Although the combinatorial calculation may include both the nuclear deformation and pairing effects to some extent, its static nature excludes a dynamical effect due to the residual interaction. 
It is well known that the missing residual interaction modifies the state density \cite{Pluhar1988, Sato1991}. 
This is especially important for strongly deformed nuclei, where rotational and vibrational collective motions enhances the transition matrix elements for the inelastic scattering process that leaves the residual nucleus in the $1p$-$1h$ configuration. 

To include this enhancement in the exciton model, we introduce a phenomenological enhancement factor into the state density as
\begin{equation}
 \omega_{\rm eff}(c,E) = K_{\rm coll}(c,E) \omega(c,E) \ ,
 \label{eq:omegacoll}
\end{equation}
where the collective enhancement factor is
\begin{equation}
 K_{\rm coll}(c,E)
 = \left\{ (\kappa - 1) \exp(-\gamma E) \right\}\delta_{n,2} + 1 \ .
 \label{eq:omega_func}
\end{equation}
The Kronecker delta on $n$ (the number of excitons), ensures $K_{\rm coll}(c,E)$ can be larger than unity when $c = (1,1,0,0)$ or $(0,0,1,1)$. 
The collective enhancement factor, $\kappa$, is an adjustable parameter ($\kappa \ge 1$) which we determine in the next section, and $\gamma$ is the damping factor such that the collective enhancement disappears at higher excitation energies. 
Because observed rotational-vibrational band heads in the nuclear structure of actinides are typically a few hundred keV or so, we empirically estimated $\gamma$ to be approximately 1~MeV. 
Note that the phenomenological enhancement factor we introduced implicitly includes a mechanism of enhanced scattering strength due to the collectiveness, which is not considered in the traditional exciton model. Hence, the calculated state density should be viewed as an effective density.

Returning to Fig.~\ref{fig:phdensity}, we see that the state density with collective enhancement, $\omega_{\rm eff}$ (dot-dashed line), shows a pronounced rise at lower excitation energy before returning to the baseline $\omega$ of Eq.~(\ref{eq:omega}) at higher excitation energy. 
The parameter $\gamma$ in Eq.~(\ref{eq:omega_func}) determines the strength of the energy dependence while $\kappa$ sets the overall scale of the enhancement. 
In the next section we determine the value of $\kappa$. 

% ========================
\section{Results} \label{sec:results}

% ------------------------
\subsection{Impact on cross sections}

Below we explore the impact of the collective enhancement for neutron-induced reactions on $^{239}$Pu. 

\begin{figure}
    \centering
    \includegraphics[width=0.49\textwidth]{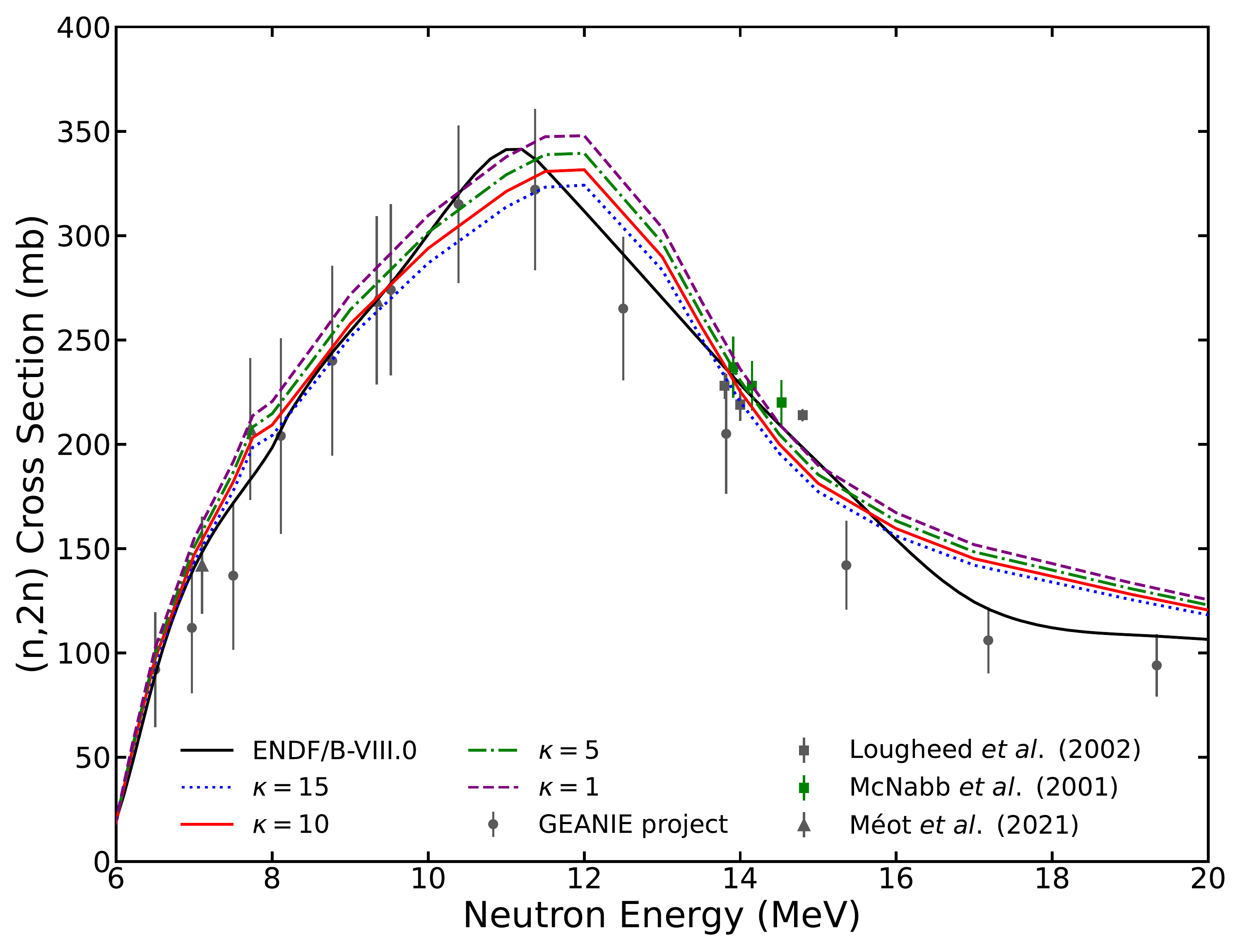}
    \caption{The $^{239}$Pu(n,2n) cross section with a range of collective enhancement factors, $\kappa = 1$, $5$, $10$, and $15$. An increased value of $\kappa$ effectively reduces the (n,2n) cross section. }
    \label{fig:n2n_cs}
\end{figure}

The influence of the collective enhancement on the $^{239}$Pu(n,2n) cross section is shown in Fig.~\ref{fig:n2n_cs}, calculated with CoH$_3$. 
As the collective enhancement factor increases from unity, the (n,2n) cross section decreases due to the shift towards a stronger pre-equilbrium reaction mechanism. 
Near threshold, there is minimal impact on the (n,2n) cross section. 
Larger differences arise starting around 8 MeV and maximizing to a spread of roughly $35$ mb around the peak of the calculated (n,2n) cross section at 12 MeV. 
The case of $\kappa=10$ performs the best with a minimal $\chi^2$ value among all datasets listed in the figure. 

\begin{figure}
    \centering
    \includegraphics[width=0.45\textwidth]{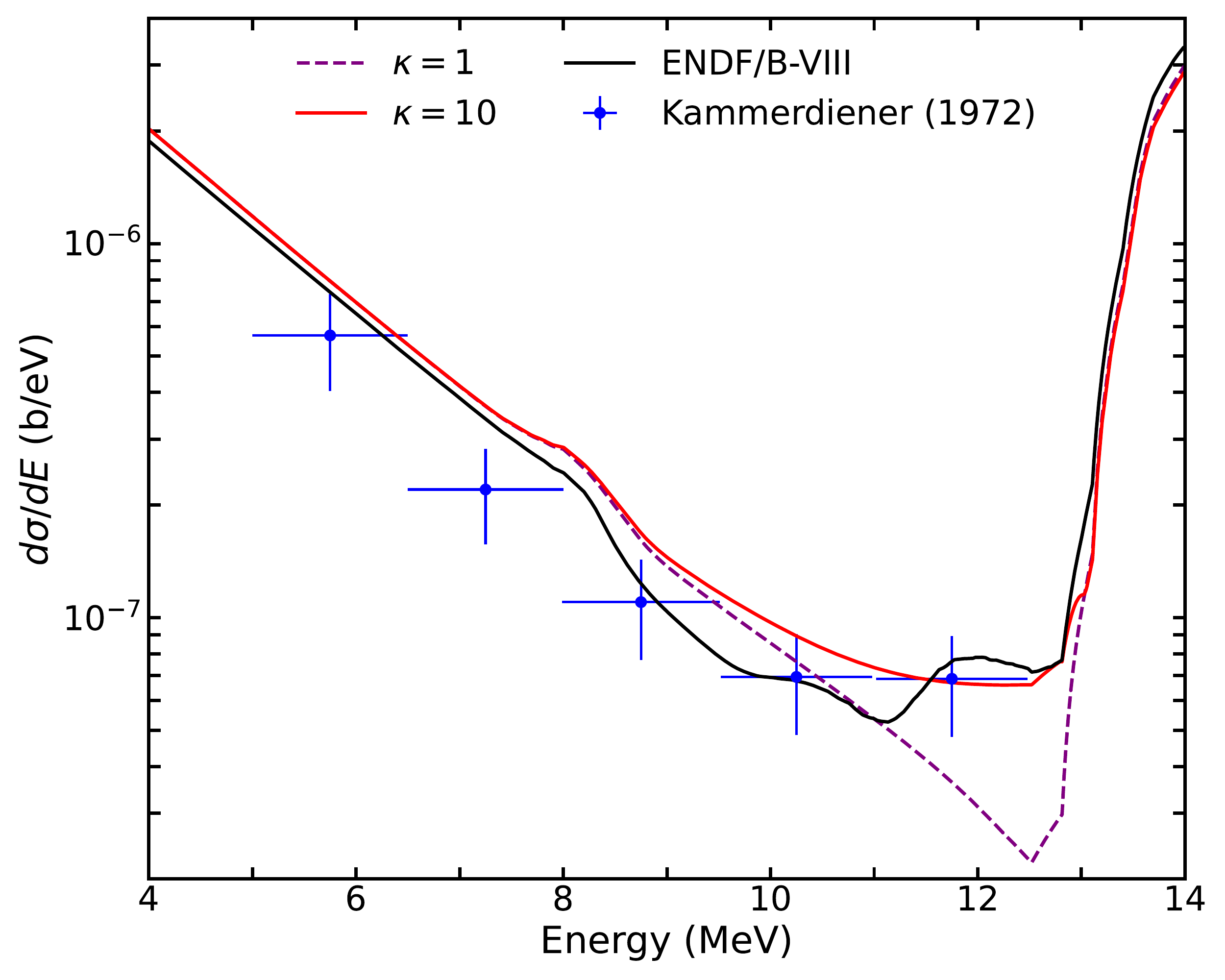}
    \caption{Neutron emission spectra for the $^{239}$Pu(n,n') reaction at incident energy of 14 MeV plotted in the outgoing neutron energy range sensitive to the collective enhancement. The present approach is compared with the standard exciton model (no collective enhancement, $\kappa=1$), ENDF/B-VIII.0 evaluation and experimental data. }
    \label{fig:DE14MeV}
\end{figure}

\begin{figure}
    \centering
    \includegraphics[width=0.45\textwidth]{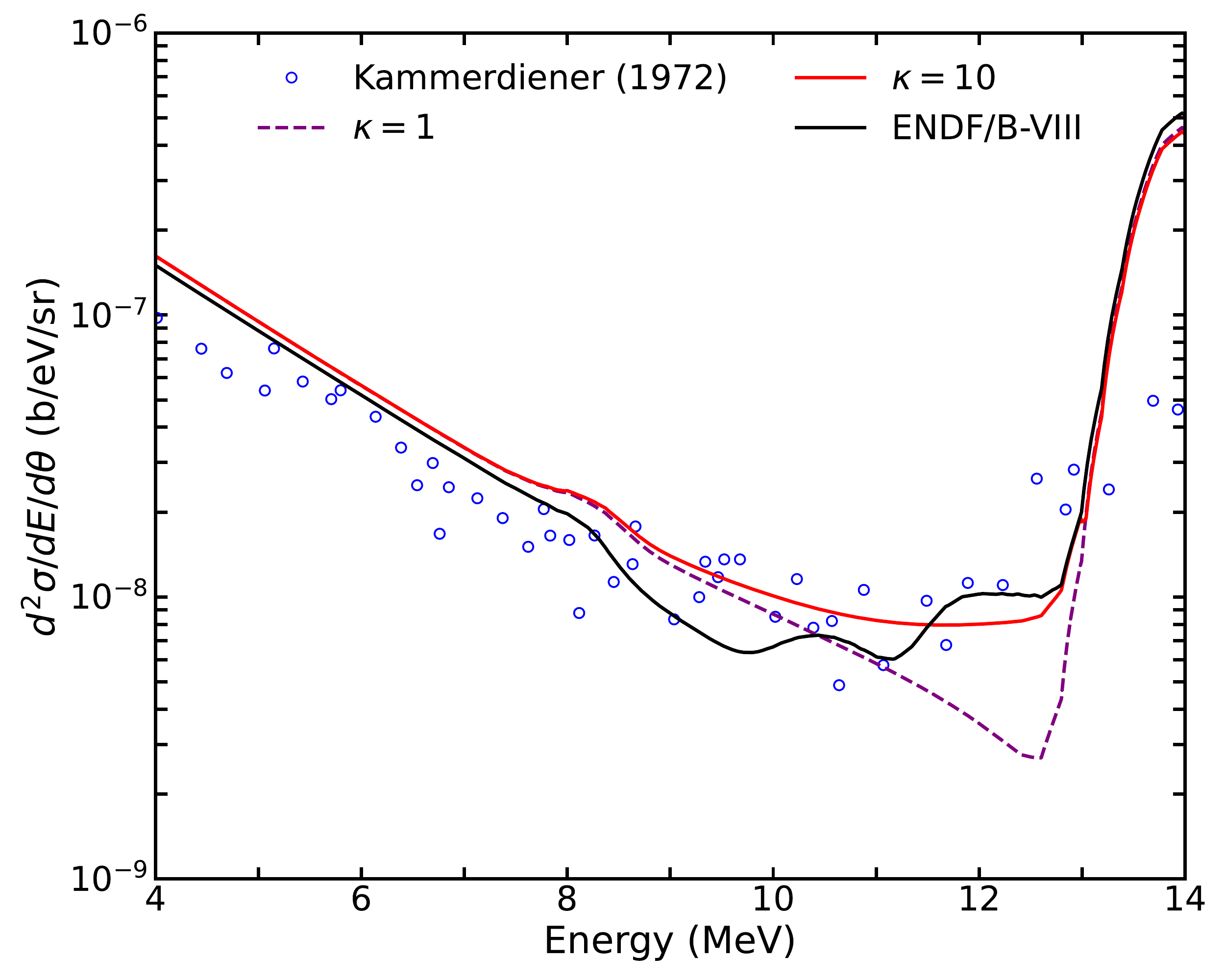} 
    \caption{As in Fig.~\ref{fig:DE14MeV} but for double-differential cross section at 40$^{\circ}$.}
    \label{fig:DDX14MeV}
\end{figure}

For inelastic scattering of 14 MeV neutrons on $^{239}$Pu, the effect of including collective enhancement in the exciton model is shown in Fig.~\ref{fig:DE14MeV} for angle integrated spectrum and in Fig.~\ref{fig:DDX14MeV} for the double-differential cross sections at 40$^{\circ}$. 
The collective enhancement causes increase of the angle integrated energy-spectrum (Fig.~\ref{fig:DE14MeV}) for outgoing neutron energies above 8 MeV. 
Around 13 MeV, this effect disappears as pre-equilibrium emission to discrete levels is turned off. 
The ENDF/B-VIII.0 results are lower than the exciton model below 11 MeV and bump up to be slightly above the present results at 12 MeV. 
This increase in the evaluated data is produced by artificial DWBA contributions to discrete levels that, for this purpose, extend up to excitation energy of 4 MeV. 
Out of the five experimental points in Fig.~\ref{fig:DE14MeV}, the first two lower energy points are below both calculations and ENDF/B-VIII.0 evaluation. 
The next two points agree better with the evaluation but are also consistent with both calculations. 
Around 12 MeV the last experimental point is well described by the present calculations and ENDF/B-VIII.0 evaluation, while the baseline exciton model is considerably lower. 
We caution that the precision of these experimental values, however, is rather low as they were obtained by averaging widely scattered double-differential data (such as those shown in Fig.~\ref{fig:DDX14MeV}). 
This reflects difficulty of measuring neutron spectra on actinides. 

% ------------------------
\subsection{Determination of collective enhancement parameter, $\kappa$}
 
The validation of the calculations against differential data and neutron spectra, which are most directly affected by the collective enhancement, turns out to be indicative but not conclusive. 
For this reason, we turn to another measured response --- the Lawrence Livermore National Laboratory (LLNL) pulsed-sphere neutron-leakage spectra. 
 
LLNL pulsed-sphere neutron-leakage spectra~\cite{PulseSpheresLLNLsummary} offer an indirect way to gauge the strength of the collective enhancement factor. 
These experiments are quasi-integral in nature: a deuteron beam hits a tritiated target in the center of a sphere mostly consisting of plutonium, and produces incident neutrons from 12--15 MeV. 
Incident neutrons scatter in the sphere material and either induce fission, releasing prompt or delayed neutrons, or scatter elastically and inelastically. 
The produced outgoing-neutron spectrum is measured at different angles as a function of time-of-flight (TOF). 
The leakage spectra are sensitive mostly to elastic and discrete inelastic levels at the earliest TOF, while the prompt-fission neutron spectrum dominates above 250 ns~\cite{Neudecker:2021,Neudecker:2021a}. 
There also is a valley in the leakage spectra right after the peak of neutrons resulting from elastic and discrete-level inelastic scattering, and before neutrons stemming from fission become dominant. 
Neutrons in this valley are produced mostly through continuum-inelastic processes. 

To determine the value of $\kappa$, we simulate LLNL pulsed-spheres neutron-leakage spectra and compare with experimental data. 
The data have reported experimental uncertainties in the range of 0.5--2\%, but are likely underestimated. 
The simulations were undertaken with the neutron-transport code MCNP-6.2~\cite{MCNP6.2}. 
All input data except those of $^{239}$Pu were taken from the latest evaluation, ENDF/B-VIII.0~\cite{ENDF8}. 

\begin{figure}
\centering
\subfigure[Neutron-leakage spectra at an angle of 117$^{\circ}$.]{\includegraphics[width=0.45\textwidth]{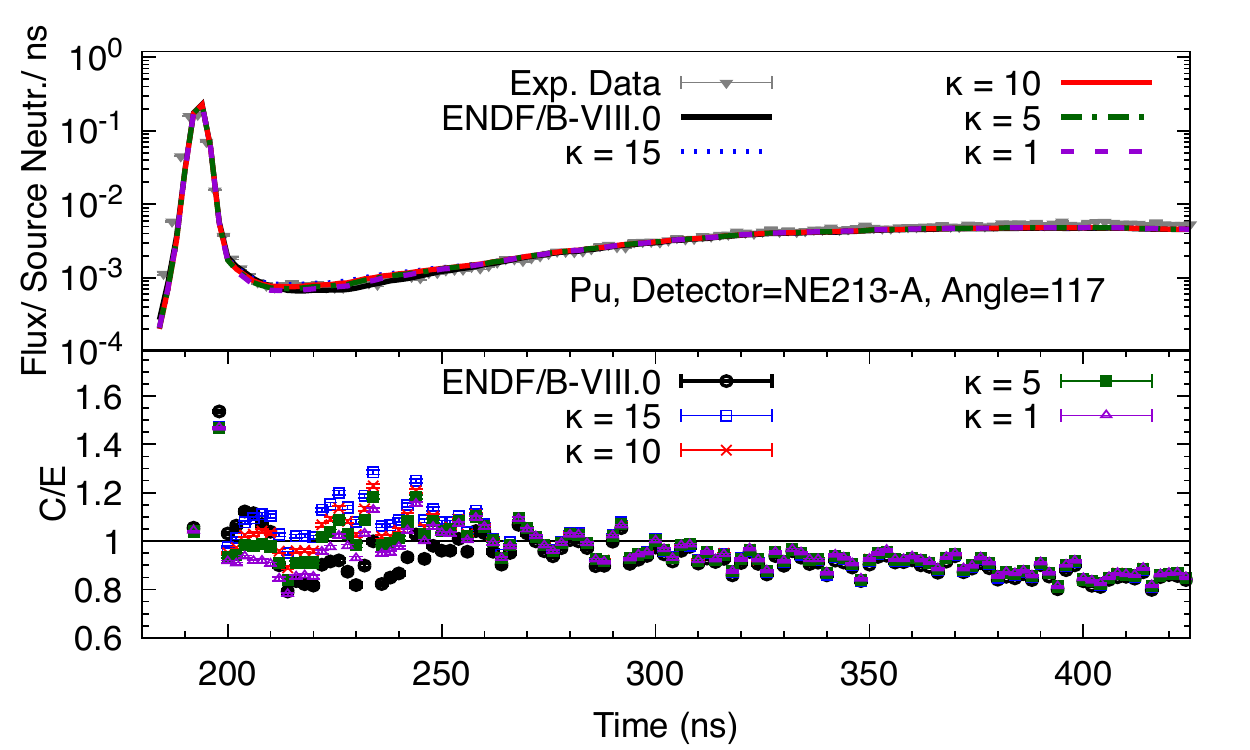}} \\
\subfigure[Neutron-leakage spectra at an angle of 26$^{\circ}$.]{\includegraphics[width=0.45\textwidth]{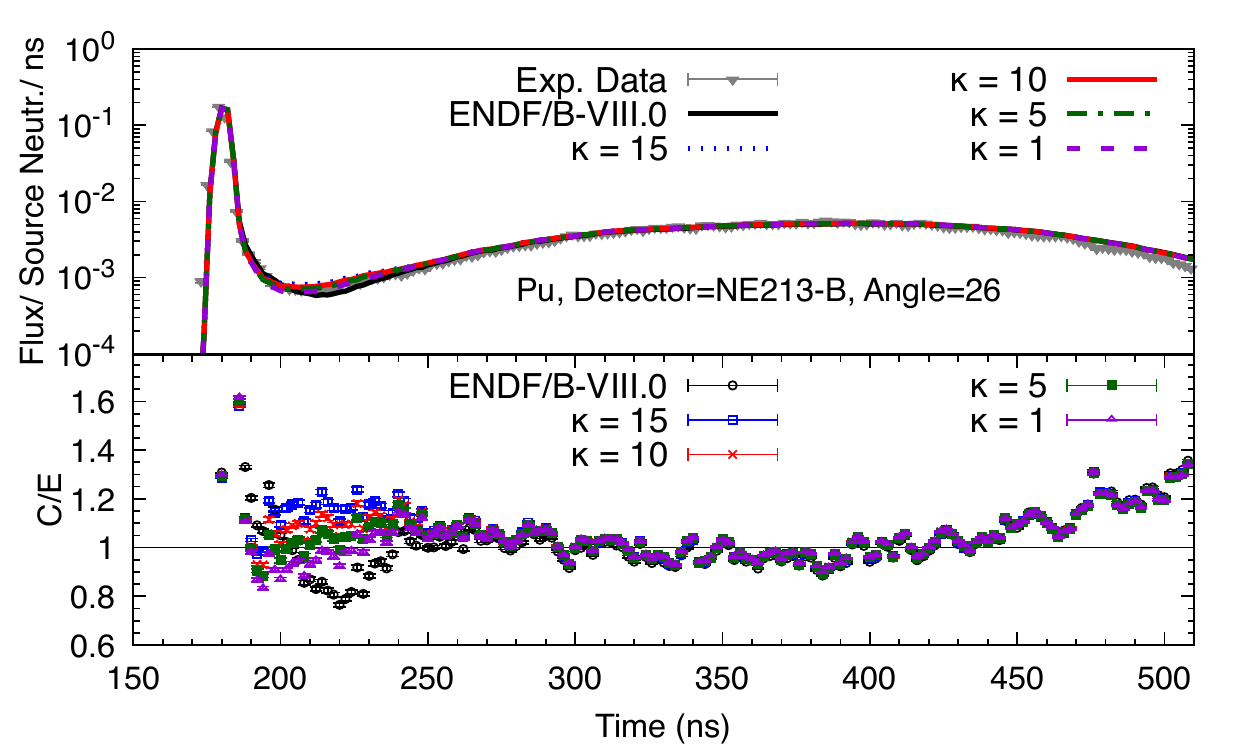}} 
\caption{LLNL pulsed-sphere neutron-leakage spectra~\cite{Neudecker:2021,PulseSpheresLLNLsummary} were simulated for a sphere of Pu with a neutron mean free path of 0.7 at two different angles with different values of the collective enhancement, $\kappa$, and are compared to ENDF/B-VIII.0 and experimental data.}
\label{fig:LLNLPS_thick}
\end{figure}

A simulation of the pulsed spheres is performed for $\kappa=1$, $5$, $10$ and $15$. 
Figure \ref{fig:LLNLPS_thick} shows the results of the simulations at different observing angles. 
From inspection of the calculated (C) to experimentally observed (E) ratios, we find that no collective enhancement ($\kappa=1$) leads to a distinct under-prediction of the neutron-leakage spectra in the valley while a collective enhancement of $\kappa=15$ leads to an over-prediction. 
Using the different angle cuts, we find that a value of $\kappa=10$ performs the best which is in agreement with the trials performed in the previous section that compared to experimental data. 

% ========================
\section{Summary}

We have proposed an enhancement to the $1p$-$1h$ state density used in the exciton model, with functional form reported in Eq.~(\ref{eq:omega_func}). 
The magnitude of the collective enhancement factor, $\kappa \approx 10$, has been estimated using comparison to LLNL pulsed spheres in conjunction with experimental data. 
The introduction of this enhancement factor allows for a better reproduction of the $^{239}$Pu (n,2n) cross section as well as double-differential cross sections in version 3 of the Los Alamos statistical Hauser-Feshbach model code, CoH. 
A consequence of the collective enhancement is the removal of ficitious DWBA levels used to simulate this effect in nuclear data evaluations \cite{ENDF8}.
The proposed modification to the $1p$-$1h$ state density thus enables a more consistent physical description of the pre-equilibrium reaction mechanism for highly-deformed actinide nuclei. 

% ========================
\begin{acknowledgments}
M.R.M. would like to acknowledge valuable discussions with Mark Chadwick and Patrick Talou. 
The authors were supported by the US Department of Energy through the Los Alamos National Laboratory (LANL). 
LANL is operated by Triad National Security, LLC, for the National Nuclear Security Administration of U.S.\ Department of Energy (Contract No.\ 89233218CNA000001). 
\end{acknowledgments}

% ========================
\bibliography{ref}

%apsrev4-2.bst 2019-01-14 (MD) hand-edited version of apsrev4-1.bst
%Control: key (0)
%Control: author (8) initials jnrlst
%Control: editor formatted (1) identically to author
%Control: production of article title (0) allowed
%Control: page (0) single
%Control: year (1) truncated
%Control: production of eprint (0) enabled
\begin{thebibliography}{36}%
\makeatletter
\providecommand \@ifxundefined [1]{%
 \@ifx{#1\undefined}
}%
\providecommand \@ifnum [1]{%
 \ifnum #1\expandafter \@firstoftwo
 \else \expandafter \@secondoftwo
 \fi
}%
\providecommand \@ifx [1]{%
 \ifx #1\expandafter \@firstoftwo
 \else \expandafter \@secondoftwo
 \fi
}%
\providecommand \natexlab [1]{#1}%
\providecommand \enquote  [1]{``#1''}%
\providecommand \bibnamefont  [1]{#1}%
\providecommand \bibfnamefont [1]{#1}%
\providecommand \citenamefont [1]{#1}%
\providecommand \href@noop [0]{\@secondoftwo}%
\providecommand \href [0]{\begingroup \@sanitize@url \@href}%
\providecommand \@href[1]{\@@startlink{#1}\@@href}%
\providecommand \@@href[1]{\endgroup#1\@@endlink}%
\providecommand \@sanitize@url [0]{\catcode `\\12\catcode `\$12\catcode
  `\&12\catcode `\#12\catcode `\^12\catcode `\_12\catcode `\%12\relax}%
\providecommand \@@startlink[1]{}%
\providecommand \@@endlink[0]{}%
\providecommand \url  [0]{\begingroup\@sanitize@url \@url }%
\providecommand \@url [1]{\endgroup\@href {#1}{\urlprefix }}%
\providecommand \urlprefix  [0]{URL }%
\providecommand \Eprint [0]{\href }%
\providecommand \doibase [0]{https://doi.org/}%
\providecommand \selectlanguage [0]{\@gobble}%
\providecommand \bibinfo  [0]{\@secondoftwo}%
\providecommand \bibfield  [0]{\@secondoftwo}%
\providecommand \translation [1]{[#1]}%
\providecommand \BibitemOpen [0]{}%
\providecommand \bibitemStop [0]{}%
\providecommand \bibitemNoStop [0]{.\EOS\space}%
\providecommand \EOS [0]{\spacefactor3000\relax}%
\providecommand \BibitemShut  [1]{\csname bibitem#1\endcsname}%
\let\auto@bib@innerbib\@empty
%</preamble>
\bibitem [{\citenamefont {Glendenning}(2004)}]{Glendenning2004}%
  \BibitemOpen
  \bibfield  {author} {\bibinfo {author} {\bibfnamefont {N.~K.}\ \bibnamefont
  {Glendenning}},\ }\href {https://doi.org/10.1142/5612} {\emph {\bibinfo
  {title} {Direct Nuclear Reactions}}}\ (\bibinfo  {publisher} {WORLD
  SCIENTIFIC},\ \bibinfo {year} {2004})\ \Eprint
  {https://arxiv.org/abs/https://www.worldscientific.com/doi/pdf/10.1142/5612}
  {https://www.worldscientific.com/doi/pdf/10.1142/5612} \BibitemShut {NoStop}%
\bibitem [{\citenamefont {Loveland}(1972)}]{Loveland1972}%
  \BibitemOpen
  \bibfield  {author} {\bibinfo {author} {\bibfnamefont {W.~D.}\ \bibnamefont
  {Loveland}},\ }\bibfield  {title} {\bibinfo {title} {Compound nuclear
  reactions},\ }\href {https://doi.org/10.1021/ed049p529} {\bibfield  {journal}
  {\bibinfo  {journal} {Journal of Chemical Education}\ }\textbf {\bibinfo
  {volume} {49}},\ \bibinfo {pages} {529} (\bibinfo {year} {1972})},\ \Eprint
  {https://arxiv.org/abs/https://doi.org/10.1021/ed049p529}
  {https://doi.org/10.1021/ed049p529} \BibitemShut {NoStop}%
\bibitem [{\citenamefont {Griffin}(1966)}]{Griffin1966}%
  \BibitemOpen
  \bibfield  {author} {\bibinfo {author} {\bibfnamefont {J.~J.}\ \bibnamefont
  {Griffin}},\ }\bibfield  {title} {\bibinfo {title} {Statistical model of
  intermediate structure},\ }\href {https://doi.org/10.1103/PhysRevLett.17.478}
  {\bibfield  {journal} {\bibinfo  {journal} {Phys. Rev. Lett.}\ }\textbf
  {\bibinfo {volume} {17}},\ \bibinfo {pages} {478} (\bibinfo {year}
  {1966})}\BibitemShut {NoStop}%
\bibitem [{\citenamefont {Gruppelaar}\ \emph {et~al.}(1986)\citenamefont
  {Gruppelaar}, \citenamefont {Nagel},\ and\ \citenamefont
  {Hodgson}}]{Gruppelaar1986}%
  \BibitemOpen
  \bibfield  {author} {\bibinfo {author} {\bibfnamefont {H.}~\bibnamefont
  {Gruppelaar}}, \bibinfo {author} {\bibfnamefont {P.}~\bibnamefont {Nagel}},\
  and\ \bibinfo {author} {\bibfnamefont {P.~E.}\ \bibnamefont {Hodgson}},\
  }\href@noop {} {\bibinfo {title} {Pre-equilibrium processes in nuclear
  reaction theory. the state-of-the-art and beyond}} (\bibinfo {year}
  {1986})\BibitemShut {NoStop}%
\bibitem [{\citenamefont {Feshbach}\ \emph {et~al.}(1980)\citenamefont
  {Feshbach}, \citenamefont {Kerman},\ and\ \citenamefont
  {Koonin}}]{Feshbach1980}%
  \BibitemOpen
  \bibfield  {author} {\bibinfo {author} {\bibfnamefont {H.}~\bibnamefont
  {Feshbach}}, \bibinfo {author} {\bibfnamefont {A.}~\bibnamefont {Kerman}},\
  and\ \bibinfo {author} {\bibfnamefont {S.}~\bibnamefont {Koonin}},\
  }\bibfield  {title} {\bibinfo {title} {The statistical theory of multi-step
  compound and direct reactions},\ }\href
  {https://doi.org/10.1016/0003-4916(80)90140-2} {\bibfield  {journal}
  {\bibinfo  {journal} {Annals of Physics}\ }\textbf {\bibinfo {volume}
  {125}},\ \bibinfo {pages} {429 } (\bibinfo {year} {1980})}\BibitemShut
  {NoStop}%
\bibitem [{\citenamefont {Tamura}\ \emph {et~al.}(1982)\citenamefont {Tamura},
  \citenamefont {Udagawa},\ and\ \citenamefont {Lenske}}]{Tamura1982}%
  \BibitemOpen
  \bibfield  {author} {\bibinfo {author} {\bibfnamefont {T.}~\bibnamefont
  {Tamura}}, \bibinfo {author} {\bibfnamefont {T.}~\bibnamefont {Udagawa}},\
  and\ \bibinfo {author} {\bibfnamefont {H.}~\bibnamefont {Lenske}},\
  }\bibfield  {title} {\bibinfo {title} {Multistep direct reaction analysis of
  continuum spectra in reactions induced by light ions},\ }\href
  {https://doi.org/10.1103/PhysRevC.26.379} {\bibfield  {journal} {\bibinfo
  {journal} {Phys. Rev. C}\ }\textbf {\bibinfo {volume} {26}},\ \bibinfo
  {pages} {379} (\bibinfo {year} {1982})}\BibitemShut {NoStop}%
\bibitem [{\citenamefont {Nishioka}\ \emph {et~al.}(1989)\citenamefont
  {Nishioka}, \citenamefont {Weidenm\"{u}ller},\ and\ \citenamefont
  {Yoshida}}]{Nishioka1989}%
  \BibitemOpen
  \bibfield  {author} {\bibinfo {author} {\bibfnamefont {H.}~\bibnamefont
  {Nishioka}}, \bibinfo {author} {\bibfnamefont {H.~A.}\ \bibnamefont
  {Weidenm\"{u}ller}},\ and\ \bibinfo {author} {\bibfnamefont {S.}~\bibnamefont
  {Yoshida}},\ }\bibfield  {title} {\bibinfo {title} {Direct-reaction effects
  in compound-nucleus and multistep compound reactions},\ }\href
  {https://doi.org/10.1016/0003-4916(89)90358-8} {\bibfield  {journal}
  {\bibinfo  {journal} {Annals of Physics}\ }\textbf {\bibinfo {volume}
  {193}},\ \bibinfo {pages} {195 } (\bibinfo {year} {1989})}\BibitemShut
  {NoStop}%
\bibitem [{\citenamefont {Luo}\ and\ \citenamefont {Kawai}(1991)}]{Luo19991}%
  \BibitemOpen
  \bibfield  {author} {\bibinfo {author} {\bibfnamefont {Y.~L.}\ \bibnamefont
  {Luo}}\ and\ \bibinfo {author} {\bibfnamefont {M.}~\bibnamefont {Kawai}},\
  }\bibfield  {title} {\bibinfo {title} {Semiclassical distorted wave model of
  nucleon inelastic scattering to continuum},\ }\href
  {https://doi.org/10.1103/PhysRevC.43.2367} {\bibfield  {journal} {\bibinfo
  {journal} {Phys. Rev. C}\ }\textbf {\bibinfo {volume} {43}},\ \bibinfo
  {pages} {2367} (\bibinfo {year} {1991})}\BibitemShut {NoStop}%
\bibitem [{\citenamefont {Kerveno}\ \emph {et~al.}(2021)\citenamefont
  {Kerveno}, \citenamefont {Dupuis}, \citenamefont {Bacquias}, \citenamefont
  {Belloni}, \citenamefont {Bernard}, \citenamefont {Borcea}, \citenamefont
  {Boromiza}, \citenamefont {Capote}, \citenamefont {De~Saint~Jean},
  \citenamefont {Dessagne}, \citenamefont {Droh\'e}, \citenamefont {Henning},
  \citenamefont {Hilaire}, \citenamefont {Kawano}, \citenamefont {Leconte},
  \citenamefont {Nankov}, \citenamefont {Negret}, \citenamefont {Nyman},
  \citenamefont {Olacel}, \citenamefont {Plompen}, \citenamefont {Romain},
  \citenamefont {Rouki}, \citenamefont {Rudolf}, \citenamefont {Stanoiu},\ and\
  \citenamefont {Wynants}}]{Kerveno2021}%
  \BibitemOpen
  \bibfield  {author} {\bibinfo {author} {\bibfnamefont {M.}~\bibnamefont
  {Kerveno}}, \bibinfo {author} {\bibfnamefont {M.}~\bibnamefont {Dupuis}},
  \bibinfo {author} {\bibfnamefont {A.}~\bibnamefont {Bacquias}}, \bibinfo
  {author} {\bibfnamefont {F.}~\bibnamefont {Belloni}}, \bibinfo {author}
  {\bibfnamefont {D.}~\bibnamefont {Bernard}}, \bibinfo {author} {\bibfnamefont
  {C.}~\bibnamefont {Borcea}}, \bibinfo {author} {\bibfnamefont
  {M.}~\bibnamefont {Boromiza}}, \bibinfo {author} {\bibfnamefont
  {R.}~\bibnamefont {Capote}}, \bibinfo {author} {\bibfnamefont
  {C.}~\bibnamefont {De~Saint~Jean}}, \bibinfo {author} {\bibfnamefont
  {P.}~\bibnamefont {Dessagne}}, \bibinfo {author} {\bibfnamefont {J.~C.}\
  \bibnamefont {Droh\'e}}, \bibinfo {author} {\bibfnamefont {G.}~\bibnamefont
  {Henning}}, \bibinfo {author} {\bibfnamefont {S.}~\bibnamefont {Hilaire}},
  \bibinfo {author} {\bibfnamefont {T.}~\bibnamefont {Kawano}}, \bibinfo
  {author} {\bibfnamefont {P.}~\bibnamefont {Leconte}}, \bibinfo {author}
  {\bibfnamefont {N.}~\bibnamefont {Nankov}}, \bibinfo {author} {\bibfnamefont
  {A.}~\bibnamefont {Negret}}, \bibinfo {author} {\bibfnamefont
  {M.}~\bibnamefont {Nyman}}, \bibinfo {author} {\bibfnamefont
  {A.}~\bibnamefont {Olacel}}, \bibinfo {author} {\bibfnamefont {A.~J.~M.}\
  \bibnamefont {Plompen}}, \bibinfo {author} {\bibfnamefont {P.}~\bibnamefont
  {Romain}}, \bibinfo {author} {\bibfnamefont {C.}~\bibnamefont {Rouki}},
  \bibinfo {author} {\bibfnamefont {G.}~\bibnamefont {Rudolf}}, \bibinfo
  {author} {\bibfnamefont {M.}~\bibnamefont {Stanoiu}},\ and\ \bibinfo {author}
  {\bibfnamefont {R.}~\bibnamefont {Wynants}},\ }\bibfield  {title} {\bibinfo
  {title} {Measurement of
  $^{238}\mathrm{U}(n,{n}^{\ensuremath{'}}\ensuremath{\gamma}$) cross section
  data and their impact on reaction models},\ }\href
  {https://doi.org/10.1103/PhysRevC.104.044605} {\bibfield  {journal} {\bibinfo
   {journal} {Phys. Rev. C}\ }\textbf {\bibinfo {volume} {104}},\ \bibinfo
  {pages} {044605} (\bibinfo {year} {2021})}\BibitemShut {NoStop}%
\bibitem [{\citenamefont {Ribanský}\ \emph {et~al.}(1973)\citenamefont
  {Ribanský}, \citenamefont {Obložinský},\ and\ \citenamefont
  {Běták}}]{Ribansky1973}%
  \BibitemOpen
  \bibfield  {author} {\bibinfo {author} {\bibfnamefont {I.}~\bibnamefont
  {Ribanský}}, \bibinfo {author} {\bibfnamefont {P.}~\bibnamefont
  {Obložinský}},\ and\ \bibinfo {author} {\bibfnamefont {E.}~\bibnamefont
  {Běták}},\ }\bibfield  {title} {\bibinfo {title} {Pre-equilibrium decay and
  the exciton model},\ }\href
  {https://doi.org/https://doi.org/10.1016/0375-9474(73)90705-7} {\bibfield
  {journal} {\bibinfo  {journal} {Nuclear Physics A}\ }\textbf {\bibinfo
  {volume} {205}},\ \bibinfo {pages} {545} (\bibinfo {year}
  {1973})}\BibitemShut {NoStop}%
\bibitem [{\citenamefont {Gudima}\ \emph {et~al.}(1983)\citenamefont {Gudima},
  \citenamefont {Mashnik},\ and\ \citenamefont {Toneev}}]{Gudima1983}%
  \BibitemOpen
  \bibfield  {author} {\bibinfo {author} {\bibfnamefont {K.}~\bibnamefont
  {Gudima}}, \bibinfo {author} {\bibfnamefont {S.}~\bibnamefont {Mashnik}},\
  and\ \bibinfo {author} {\bibfnamefont {V.}~\bibnamefont {Toneev}},\
  }\bibfield  {title} {\bibinfo {title} {Cascade-exciton model of nuclear
  reactions},\ }\href
  {https://doi.org/https://doi.org/10.1016/0375-9474(83)90532-8} {\bibfield
  {journal} {\bibinfo  {journal} {Nuclear Physics A}\ }\textbf {\bibinfo
  {volume} {401}},\ \bibinfo {pages} {329} (\bibinfo {year}
  {1983})}\BibitemShut {NoStop}%
\bibitem [{\citenamefont {Kawano}\ \emph {et~al.}(2006)\citenamefont {Kawano},
  \citenamefont {Talou},\ and\ \citenamefont {Chadwick}}]{Kawano2006b}%
  \BibitemOpen
  \bibfield  {author} {\bibinfo {author} {\bibfnamefont {T.}~\bibnamefont
  {Kawano}}, \bibinfo {author} {\bibfnamefont {P.}~\bibnamefont {Talou}},\ and\
  \bibinfo {author} {\bibfnamefont {M.~B.}\ \bibnamefont {Chadwick}},\
  }\bibfield  {title} {\bibinfo {title} {Production of isomers by
  neutron-induced inelastic scattering on $^{193}$ir and influence of spin
  distribution in the pre-equilibrium process},\ }\href
  {https://doi.org/https://doi.org/10.1016/j.nima.2006.02.053} {\bibfield
  {journal} {\bibinfo  {journal} {Nuclear Instruments and Methods in Physics
  Research Section A: Accelerators, Spectrometers, Detectors and Associated
  Equipment}\ }\textbf {\bibinfo {volume} {562}},\ \bibinfo {pages} {774 }
  (\bibinfo {year} {2006})},\ \bibinfo {note} {proceedings of the 7th
  International Conference on Accelerator Applications}\BibitemShut {NoStop}%
\bibitem [{\citenamefont {Dashdorj}\ \emph {et~al.}(2007)\citenamefont
  {Dashdorj}, \citenamefont {Kawano}, \citenamefont {Garrett}, \citenamefont
  {Becker}, \citenamefont {Agvaanluvsan}, \citenamefont {Bernstein},
  \citenamefont {Chadwick}, \citenamefont {Devlin}, \citenamefont {Fotiades},
  \citenamefont {Mitchell}, \citenamefont {Nelson},\ and\ \citenamefont
  {Younes}}]{Dashdorj2007}%
  \BibitemOpen
  \bibfield  {author} {\bibinfo {author} {\bibfnamefont {D.}~\bibnamefont
  {Dashdorj}}, \bibinfo {author} {\bibfnamefont {T.}~\bibnamefont {Kawano}},
  \bibinfo {author} {\bibfnamefont {P.~E.}\ \bibnamefont {Garrett}}, \bibinfo
  {author} {\bibfnamefont {J.~A.}\ \bibnamefont {Becker}}, \bibinfo {author}
  {\bibfnamefont {U.}~\bibnamefont {Agvaanluvsan}}, \bibinfo {author}
  {\bibfnamefont {L.~A.}\ \bibnamefont {Bernstein}}, \bibinfo {author}
  {\bibfnamefont {M.~B.}\ \bibnamefont {Chadwick}}, \bibinfo {author}
  {\bibfnamefont {M.}~\bibnamefont {Devlin}}, \bibinfo {author} {\bibfnamefont
  {N.}~\bibnamefont {Fotiades}}, \bibinfo {author} {\bibfnamefont {G.~E.}\
  \bibnamefont {Mitchell}}, \bibinfo {author} {\bibfnamefont {R.~O.}\
  \bibnamefont {Nelson}},\ and\ \bibinfo {author} {\bibfnamefont
  {W.}~\bibnamefont {Younes}},\ }\bibfield  {title} {\bibinfo {title} {{Effect
  of preequilibrium spin distribution on $^{48}\mathrm{Ti}$ $+n$ cross
  sections}},\ }\href {https://doi.org/10.1103/PhysRevC.75.054612} {\bibfield
  {journal} {\bibinfo  {journal} {Phys. Rev. C}\ }\textbf {\bibinfo {volume}
  {75}},\ \bibinfo {pages} {054612} (\bibinfo {year} {2007})}\BibitemShut
  {NoStop}%
\bibitem [{\citenamefont {Dupuis}\ \emph {et~al.}(2015)\citenamefont {Dupuis},
  \citenamefont {Bauge}, \citenamefont {Hilaire}, \citenamefont {Lechaftois},
  \citenamefont {P\'{e}ru}, \citenamefont {Pillet},\ and\ \citenamefont
  {Robin}}]{Dupuis2015}%
  \BibitemOpen
  \bibfield  {author} {\bibinfo {author} {\bibfnamefont {M.}~\bibnamefont
  {Dupuis}}, \bibinfo {author} {\bibfnamefont {E.}~\bibnamefont {Bauge}},
  \bibinfo {author} {\bibfnamefont {S.}~\bibnamefont {Hilaire}}, \bibinfo
  {author} {\bibfnamefont {F.}~\bibnamefont {Lechaftois}}, \bibinfo {author}
  {\bibfnamefont {S.}~\bibnamefont {P\'{e}ru}}, \bibinfo {author}
  {\bibfnamefont {N.}~\bibnamefont {Pillet}},\ and\ \bibinfo {author}
  {\bibfnamefont {C.}~\bibnamefont {Robin}},\ }\bibfield  {title} {\bibinfo
  {title} {Progress in microscopic direct reaction modeling of nucleon induced
  reactions},\ }\href {https://doi.org/10.1140/epja/i2015-15168-x} {\bibfield
  {journal} {\bibinfo  {journal} {European Physical Journal A}\ ,\ \bibinfo
  {pages} {168}} (\bibinfo {year} {2015})}\BibitemShut {NoStop}%
\bibitem [{\citenamefont {{Dupuis, M.}}\ \emph {et~al.}(2017)\citenamefont
  {{Dupuis, M.}}, \citenamefont {{Hilaire, S.}}, \citenamefont {{P\'eru, S.}},
  \citenamefont {{Bauge, E.}}, \citenamefont {{Kerveno, M.}}, \citenamefont
  {{Dessagne, P.}},\ and\ \citenamefont {{Henning, G.}}}]{Dupuis2017}%
  \BibitemOpen
  \bibfield  {author} {\bibinfo {author} {\bibnamefont {{Dupuis, M.}}},
  \bibinfo {author} {\bibnamefont {{Hilaire, S.}}}, \bibinfo {author}
  {\bibnamefont {{P\'eru, S.}}}, \bibinfo {author} {\bibnamefont {{Bauge,
  E.}}}, \bibinfo {author} {\bibnamefont {{Kerveno, M.}}}, \bibinfo {author}
  {\bibnamefont {{Dessagne, P.}}},\ and\ \bibinfo {author} {\bibnamefont
  {{Henning, G.}}},\ }\bibfield  {title} {\bibinfo {title} {Microscopic
  modeling of direct pre-equilibrium emission from neutron induced reactions on
  even and odd actinides},\ }\href
  {https://doi.org/10.1051/epjconf/201714612002} {\bibfield  {journal}
  {\bibinfo  {journal} {EPJ Web Conf.}\ }\textbf {\bibinfo {volume} {146}},\
  \bibinfo {pages} {12002} (\bibinfo {year} {2017})}\BibitemShut {NoStop}%
\bibitem [{\citenamefont {Kawano}\ \emph {et~al.}(2016)\citenamefont {Kawano},
  \citenamefont {Capote}, \citenamefont {Hilaire},\ and\ \citenamefont {Chau
  Huu-Tai}}]{Kawano2016}%
  \BibitemOpen
  \bibfield  {author} {\bibinfo {author} {\bibfnamefont {T.}~\bibnamefont
  {Kawano}}, \bibinfo {author} {\bibfnamefont {R.}~\bibnamefont {Capote}},
  \bibinfo {author} {\bibfnamefont {S.}~\bibnamefont {Hilaire}},\ and\ \bibinfo
  {author} {\bibfnamefont {P.}~\bibnamefont {Chau Huu-Tai}},\ }\bibfield
  {title} {\bibinfo {title} {Statistical hauser-feshbach theory with
  width-fluctuation correction including direct reaction channels for
  neutron-induced reactions at low energies},\ }\href
  {https://doi.org/10.1103/PhysRevC.94.014612} {\bibfield  {journal} {\bibinfo
  {journal} {Phys. Rev. C}\ }\textbf {\bibinfo {volume} {94}},\ \bibinfo
  {pages} {014612} (\bibinfo {year} {2016})}\BibitemShut {NoStop}%
\bibitem [{\citenamefont {Kawano}(2019)}]{Kawano2019}%
  \BibitemOpen
  \bibfield  {author} {\bibinfo {author} {\bibfnamefont {T.}~\bibnamefont
  {Kawano}},\ }\href {https://doi.org/10.48550/ARXIV.1901.05641} {\bibinfo
  {title} {Unified coupled-channels and hauser-feshbach model calculation for
  nuclear data evaluation}} (\bibinfo {year} {2019})\BibitemShut {NoStop}%
\bibitem [{\citenamefont {Kalbach}(1986)}]{Kalbach1986}%
  \BibitemOpen
  \bibfield  {author} {\bibinfo {author} {\bibfnamefont {C.}~\bibnamefont
  {Kalbach}},\ }\bibfield  {title} {\bibinfo {title} {Two-component exciton
  model: Basic formalism away from shell closures},\ }\href
  {https://doi.org/10.1103/PhysRevC.33.818} {\bibfield  {journal} {\bibinfo
  {journal} {Phys. Rev. C}\ }\textbf {\bibinfo {volume} {33}},\ \bibinfo
  {pages} {818} (\bibinfo {year} {1986})}\BibitemShut {NoStop}%
\bibitem [{\citenamefont {Koning}\ and\ \citenamefont
  {Duijvestijn}(2004)}]{Koning2004}%
  \BibitemOpen
  \bibfield  {author} {\bibinfo {author} {\bibfnamefont {A.~J.}\ \bibnamefont
  {Koning}}\ and\ \bibinfo {author} {\bibfnamefont {M.~C.}\ \bibnamefont
  {Duijvestijn}},\ }\bibfield  {title} {\bibinfo {title} {A global
  pre-equilibrium analysis from 7 to 200 mev based on the optical model
  potential},\ }\href {https://doi.org/10.1016/j.nuclphysa.2004.08.013}
  {\bibfield  {journal} {\bibinfo  {journal} {Nuclear Physics A}\ }\textbf
  {\bibinfo {volume} {744}},\ \bibinfo {pages} {15 } (\bibinfo {year}
  {2004})}\BibitemShut {NoStop}%
\bibitem [{\citenamefont {B{\v{e}}t{\'a}k}\ and\ \citenamefont
  {Dobe{\v{s}}}(1976)}]{Betak1976}%
  \BibitemOpen
  \bibfield  {author} {\bibinfo {author} {\bibfnamefont {E.}~\bibnamefont
  {B{\v{e}}t{\'a}k}}\ and\ \bibinfo {author} {\bibfnamefont {J.}~\bibnamefont
  {Dobe{\v{s}}}},\ }\bibfield  {title} {\bibinfo {title} {The finite depth of
  the nuclear potential well in the exciton model of preequilibrium decay},\
  }\href {https://doi.org/10.1007/BF01408305} {\bibfield  {journal} {\bibinfo
  {journal} {Zeitschrift f{\"u}r Physik A Atoms and Nuclei}\ }\textbf {\bibinfo
  {volume} {279}},\ \bibinfo {pages} {319 } (\bibinfo {year}
  {1976})}\BibitemShut {NoStop}%
\bibitem [{\citenamefont {Kalbach}(2001)}]{Kalbach2006}%
  \BibitemOpen
  \bibfield  {author} {\bibinfo {author} {\bibfnamefont {C.}~\bibnamefont
  {Kalbach}},\ }\href@noop {} {\bibinfo {title} {Users manual for preco-2006}}
  (\bibinfo {year} {2001})\BibitemShut {NoStop}%
\bibitem [{\citenamefont {Williams}(1971)}]{Williams1971}%
  \BibitemOpen
  \bibfield  {author} {\bibinfo {author} {\bibfnamefont {F.~C.}\ \bibnamefont
  {Williams}},\ }\bibfield  {title} {\bibinfo {title} {Particle-hole state
  density in the uniform spacing model},\ }\href
  {https://doi.org/10.1016/0375-9474(71)90426-X} {\bibfield  {journal}
  {\bibinfo  {journal} {Nuclear Physics A}\ }\textbf {\bibinfo {volume}
  {166}},\ \bibinfo {pages} {231 } (\bibinfo {year} {1971})}\BibitemShut
  {NoStop}%
\bibitem [{\citenamefont {Fu}(1980)}]{ORNL-7042}%
  \BibitemOpen
  \bibfield  {author} {\bibinfo {author} {\bibfnamefont {C.~Y.}\ \bibnamefont
  {Fu}},\ }\href@noop {} {\emph {\bibinfo {title} {A Consistent Nuclear Model
  for Compound and Precompound Reactions with Conservation of Angular
  Momentum}}},\ \bibinfo {type} {Tech. Rep.}\ \bibinfo {number} {ORNL/TM-7042}\
  (\bibinfo  {institution} {Oak Ridge National Laboratory},\ \bibinfo {year}
  {1980})\BibitemShut {NoStop}%
\bibitem [{\citenamefont {Strutinsky}(1968)}]{Strutinsky1968}%
  \BibitemOpen
  \bibfield  {author} {\bibinfo {author} {\bibfnamefont {V.~M.}\ \bibnamefont
  {Strutinsky}},\ }\bibfield  {title} {\bibinfo {title} {"shells" in deformed
  nuclei},\ }\href {https://doi.org/10.1016/0375-9474(68)90699-4} {\bibfield
  {journal} {\bibinfo  {journal} {Nuclear Physics A}\ }\textbf {\bibinfo
  {volume} {122}},\ \bibinfo {pages} {1 } (\bibinfo {year} {1968})}\BibitemShut
  {NoStop}%
\bibitem [{\citenamefont {Bolsterli}\ \emph {et~al.}(1972)\citenamefont
  {Bolsterli}, \citenamefont {Fiset}, \citenamefont {Nix},\ and\ \citenamefont
  {Norton}}]{Bolsterli1972}%
  \BibitemOpen
  \bibfield  {author} {\bibinfo {author} {\bibfnamefont {M.}~\bibnamefont
  {Bolsterli}}, \bibinfo {author} {\bibfnamefont {E.~O.}\ \bibnamefont
  {Fiset}}, \bibinfo {author} {\bibfnamefont {J.~R.}\ \bibnamefont {Nix}},\
  and\ \bibinfo {author} {\bibfnamefont {J.~L.}\ \bibnamefont {Norton}},\
  }\bibfield  {title} {\bibinfo {title} {New calculation of fission barriers
  for heavy and superheavy nuclei},\ }\href
  {https://doi.org/10.1103/PhysRevC.5.1050} {\bibfield  {journal} {\bibinfo
  {journal} {Phys. Rev. C}\ }\textbf {\bibinfo {volume} {5}},\ \bibinfo {pages}
  {1050 } (\bibinfo {year} {1972})}\BibitemShut {NoStop}%
\bibitem [{\citenamefont {Shlomo}(1992)}]{Shlomo1991}%
  \BibitemOpen
  \bibfield  {author} {\bibinfo {author} {\bibfnamefont {S.}~\bibnamefont
  {Shlomo}},\ }\bibfield  {title} {\bibinfo {title} {Energy level density of
  nuclei},\ }\href {https://doi.org/10.1016/0375-9474(92)90233-A} {\bibfield
  {journal} {\bibinfo  {journal} {Nuclear Physics A}\ }\textbf {\bibinfo
  {volume} {539}},\ \bibinfo {pages} {17 } (\bibinfo {year}
  {1992})}\BibitemShut {NoStop}%
\bibitem [{\citenamefont {M\"{o}ller}\ \emph {et~al.}(1995)\citenamefont
  {M\"{o}ller}, \citenamefont {Nix}, \citenamefont {Myer},\ and\ \citenamefont
  {Swiatecki}}]{Moller1995}%
  \BibitemOpen
  \bibfield  {author} {\bibinfo {author} {\bibfnamefont {P.}~\bibnamefont
  {M\"{o}ller}}, \bibinfo {author} {\bibfnamefont {J.~R.}\ \bibnamefont {Nix}},
  \bibinfo {author} {\bibfnamefont {W.~D.}\ \bibnamefont {Myer}},\ and\
  \bibinfo {author} {\bibfnamefont {W.~J.}\ \bibnamefont {Swiatecki}},\
  }\bibfield  {title} {\bibinfo {title} {Nuclear ground-state masses and
  deformations},\ }\href {https://doi.org/10.1006/adnd.1995.1002} {\bibfield
  {journal} {\bibinfo  {journal} {Atomic Data and Nuclear Data Tables}\
  }\textbf {\bibinfo {volume} {59}},\ \bibinfo {pages} {185 } (\bibinfo {year}
  {1995})}\BibitemShut {NoStop}%
\bibitem [{\citenamefont {M\"{o}ller}\ \emph {et~al.}(2016)\citenamefont
  {M\"{o}ller}, \citenamefont {Sierk}, \citenamefont {Ichikawa},\ and\
  \citenamefont {Sagawa}}]{Moller2016}%
  \BibitemOpen
  \bibfield  {author} {\bibinfo {author} {\bibfnamefont {P.}~\bibnamefont
  {M\"{o}ller}}, \bibinfo {author} {\bibfnamefont {A.~J.}\ \bibnamefont
  {Sierk}}, \bibinfo {author} {\bibfnamefont {T.}~\bibnamefont {Ichikawa}},\
  and\ \bibinfo {author} {\bibfnamefont {H.}~\bibnamefont {Sagawa}},\
  }\bibfield  {title} {\bibinfo {title} {Nuclear ground-state masses and
  deformations: Frdm(2012)},\ }\href
  {https://doi.org/10.1016/j.adt.2015.10.002} {\bibfield  {journal} {\bibinfo
  {journal} {Atomic Data and Nuclear Data Tables}\ }\textbf {\bibinfo {volume}
  {109 -- 110}},\ \bibinfo {pages} {1 } (\bibinfo {year} {2016})}\BibitemShut
  {NoStop}%
\bibitem [{\citenamefont {Herman}\ and\ \citenamefont
  {Reffo}(1987)}]{Herman1987}%
  \BibitemOpen
  \bibfield  {author} {\bibinfo {author} {\bibfnamefont {M.}~\bibnamefont
  {Herman}}\ and\ \bibinfo {author} {\bibfnamefont {G.}~\bibnamefont {Reffo}},\
  }\bibfield  {title} {\bibinfo {title} {Realistic few-quasiparticle level
  densities in spherical nuclei},\ }\href
  {https://doi.org/10.1103/PhysRevC.36.1546} {\bibfield  {journal} {\bibinfo
  {journal} {Phys. Rev. C}\ }\textbf {\bibinfo {volume} {36}},\ \bibinfo
  {pages} {1546} (\bibinfo {year} {1987})}\BibitemShut {NoStop}%
\bibitem [{\citenamefont {Pluha\v{r}}\ and\ \citenamefont
  {Weidenm\"uller}(1988)}]{Pluhar1988}%
  \BibitemOpen
  \bibfield  {author} {\bibinfo {author} {\bibfnamefont {Z.}~\bibnamefont
  {Pluha\v{r}}}\ and\ \bibinfo {author} {\bibfnamefont {H.~A.}\ \bibnamefont
  {Weidenm\"uller}},\ }\bibfield  {title} {\bibinfo {title} {Approximation for
  shell-model level densities},\ }\href
  {https://doi.org/10.1103/PhysRevC.38.1046} {\bibfield  {journal} {\bibinfo
  {journal} {Phys. Rev. C}\ }\textbf {\bibinfo {volume} {38}},\ \bibinfo
  {pages} {1046 } (\bibinfo {year} {1988})}\BibitemShut {NoStop}%
\bibitem [{\citenamefont {Sato}\ \emph {et~al.}(1991)\citenamefont {Sato},
  \citenamefont {Takahashi},\ and\ \citenamefont {Yoshida}}]{Sato1991}%
  \BibitemOpen
  \bibfield  {author} {\bibinfo {author} {\bibfnamefont {K.}~\bibnamefont
  {Sato}}, \bibinfo {author} {\bibfnamefont {Y.}~\bibnamefont {Takahashi}},\
  and\ \bibinfo {author} {\bibfnamefont {S.}~\bibnamefont {Yoshida}},\
  }\bibfield  {title} {\bibinfo {title} {Exciton level densities with spin and
  parity based on random matrix model},\ }\href
  {https://doi.org/10.1007/BF01282942} {\bibfield  {journal} {\bibinfo
  {journal} {Zeitschrift f{\"u}r Physik A Hadrons and Nuclei}\ }\textbf
  {\bibinfo {volume} {339}},\ \bibinfo {pages} {129 } (\bibinfo {year}
  {1991})}\BibitemShut {NoStop}%
\bibitem [{\citenamefont {Wong}\ \emph {et~al.}(1972)\citenamefont {Wong},
  \citenamefont {Anderson}, \citenamefont {Brown}, \citenamefont {Hansen},
  \citenamefont {Kammerdiener}, \citenamefont {Logan},\ and\ \citenamefont
  {Pohl}}]{PulseSpheresLLNLsummary}%
  \BibitemOpen
  \bibfield  {author} {\bibinfo {author} {\bibfnamefont {C.}~\bibnamefont
  {Wong}}, \bibinfo {author} {\bibfnamefont {J.}~\bibnamefont {Anderson}},
  \bibinfo {author} {\bibfnamefont {P.}~\bibnamefont {Brown}}, \bibinfo
  {author} {\bibfnamefont {L.~F.}\ \bibnamefont {Hansen}}, \bibinfo {author}
  {\bibfnamefont {J.~L.}\ \bibnamefont {Kammerdiener}}, \bibinfo {author}
  {\bibfnamefont {C.}~\bibnamefont {Logan}},\ and\ \bibinfo {author}
  {\bibfnamefont {B.}~\bibnamefont {Pohl}},\ }\bibfield  {title} {\bibinfo
  {title} {Livermore pulsed sphere program: Program summary through july
  1971},\ }\href@noop {} {\bibfield  {journal} {\bibinfo  {journal} {LLNL
  UCRL-51144 Rev. 1}\ } (\bibinfo {year} {1972})}\BibitemShut {NoStop}%
\bibitem [{\citenamefont {Neudecker}\ \emph
  {et~al.}(2021{\natexlab{a}})\citenamefont {Neudecker}, \citenamefont
  {Cabellos}, \citenamefont {Clark}, \citenamefont {Haeck}, \citenamefont
  {Capote}, \citenamefont {Trkov}, \citenamefont {White},\ and\ \citenamefont
  {Rising}}]{Neudecker:2021}%
  \BibitemOpen
  \bibfield  {author} {\bibinfo {author} {\bibfnamefont {D.}~\bibnamefont
  {Neudecker}}, \bibinfo {author} {\bibfnamefont {O.}~\bibnamefont {Cabellos}},
  \bibinfo {author} {\bibfnamefont {A.}~\bibnamefont {Clark}}, \bibinfo
  {author} {\bibfnamefont {W.}~\bibnamefont {Haeck}}, \bibinfo {author}
  {\bibfnamefont {R.}~\bibnamefont {Capote}}, \bibinfo {author} {\bibfnamefont
  {A.}~\bibnamefont {Trkov}}, \bibinfo {author} {\bibfnamefont {M.~C.}\
  \bibnamefont {White}},\ and\ \bibinfo {author} {\bibfnamefont {M.~E.}\
  \bibnamefont {Rising}},\ }\bibfield  {title} {\bibinfo {title} {Which nuclear
  data can be validated with llnl pulsed-sphere experiments?},\ }\href
  {https://doi.org/https://doi.org/10.1016/j.anucene.2021.108345} {\bibfield
  {journal} {\bibinfo  {journal} {Ann. Nucl. Energy}\ }\textbf {\bibinfo
  {volume} {159}},\ \bibinfo {pages} {108345} (\bibinfo {year}
  {2021}{\natexlab{a}})}\BibitemShut {NoStop}%
\bibitem [{\citenamefont {Neudecker}\ \emph
  {et~al.}(2021{\natexlab{b}})\citenamefont {Neudecker}, \citenamefont
  {Cabellos}, \citenamefont {Clark}, \citenamefont {Grosskopf}, \citenamefont
  {Haeck}, \citenamefont {Herman}, \citenamefont {Hutchinson}, \citenamefont
  {Kawano}, \citenamefont {Lovell}, \citenamefont {Stetcu}, \citenamefont
  {Talou},\ and\ \citenamefont {Vander~Wiel}}]{Neudecker:2021a}%
  \BibitemOpen
  \bibfield  {author} {\bibinfo {author} {\bibfnamefont {D.}~\bibnamefont
  {Neudecker}}, \bibinfo {author} {\bibfnamefont {O.}~\bibnamefont {Cabellos}},
  \bibinfo {author} {\bibfnamefont {A.~R.}\ \bibnamefont {Clark}}, \bibinfo
  {author} {\bibfnamefont {M.~J.}\ \bibnamefont {Grosskopf}}, \bibinfo {author}
  {\bibfnamefont {W.}~\bibnamefont {Haeck}}, \bibinfo {author} {\bibfnamefont
  {M.~W.}\ \bibnamefont {Herman}}, \bibinfo {author} {\bibfnamefont
  {J.}~\bibnamefont {Hutchinson}}, \bibinfo {author} {\bibfnamefont
  {T.}~\bibnamefont {Kawano}}, \bibinfo {author} {\bibfnamefont {A.~E.}\
  \bibnamefont {Lovell}}, \bibinfo {author} {\bibfnamefont {I.}~\bibnamefont
  {Stetcu}}, \bibinfo {author} {\bibfnamefont {P.}~\bibnamefont {Talou}},\ and\
  \bibinfo {author} {\bibfnamefont {S.}~\bibnamefont {Vander~Wiel}},\
  }\bibfield  {title} {\bibinfo {title} {Informing nuclear physics via machine
  learning methods with differential and integral experiments},\ }\href
  {https://doi.org/10.1103/PhysRevC.104.034611} {\bibfield  {journal} {\bibinfo
   {journal} {Phys. Rev. C}\ }\textbf {\bibinfo {volume} {104}},\ \bibinfo
  {pages} {034611} (\bibinfo {year} {2021}{\natexlab{b}})}\BibitemShut
  {NoStop}%
\bibitem [{\citenamefont {Werner}\ \emph {et~al.}(2017)\citenamefont {Werner},
  \citenamefont {Armstrong}, \citenamefont {Brown}, \citenamefont {Bull},
  \citenamefont {Casswell}, \citenamefont {Cox}, \citenamefont {Dixon},
  \citenamefont {Forster}, \citenamefont {Goorley}, \citenamefont {Hughes},
  \citenamefont {Favorite}, \citenamefont {Martz}, \citenamefont {Mashnik},
  \citenamefont {Rising}, \citenamefont {Solomon}, \citenamefont {Sood},
  \citenamefont {Sweezy}, \citenamefont {Zukaitis}, \citenamefont {Anderson},
  \citenamefont {Elson}, \citenamefont {Durkee}, \citenamefont {Johns},
  \citenamefont {McKinney}, \citenamefont {McMath}, \citenamefont {Hendricks},
  \citenamefont {Pelowitz}, \citenamefont {Prael}, \citenamefont {Booth},
  \citenamefont {James}, \citenamefont {Fensin}, \citenamefont {Wilcox},\ and\
  \citenamefont {Kiedrowski}}]{MCNP6.2}%
  \BibitemOpen
  \bibfield  {author} {\bibinfo {author} {\bibfnamefont {C.}~\bibnamefont
  {Werner}}, \bibinfo {author} {\bibfnamefont {J.}~\bibnamefont {Armstrong}},
  \bibinfo {author} {\bibfnamefont {F.}~\bibnamefont {Brown}}, \bibinfo
  {author} {\bibfnamefont {J.}~\bibnamefont {Bull}}, \bibinfo {author}
  {\bibfnamefont {L.}~\bibnamefont {Casswell}}, \bibinfo {author}
  {\bibfnamefont {L.}~\bibnamefont {Cox}}, \bibinfo {author} {\bibfnamefont
  {D.}~\bibnamefont {Dixon}}, \bibinfo {author} {\bibfnamefont
  {R.}~\bibnamefont {Forster}}, \bibinfo {author} {\bibfnamefont
  {J.}~\bibnamefont {Goorley}}, \bibinfo {author} {\bibfnamefont
  {H.}~\bibnamefont {Hughes}}, \bibinfo {author} {\bibfnamefont
  {J.}~\bibnamefont {Favorite}}, \bibinfo {author} {\bibfnamefont
  {R.}~\bibnamefont {Martz}}, \bibinfo {author} {\bibfnamefont
  {S.}~\bibnamefont {Mashnik}}, \bibinfo {author} {\bibfnamefont
  {M.}~\bibnamefont {Rising}}, \bibinfo {author} {\bibfnamefont
  {C.}~\bibnamefont {Solomon}}, \bibinfo {author} {\bibfnamefont
  {A.}~\bibnamefont {Sood}}, \bibinfo {author} {\bibfnamefont {J.}~\bibnamefont
  {Sweezy}}, \bibinfo {author} {\bibfnamefont {A.}~\bibnamefont {Zukaitis}},
  \bibinfo {author} {\bibfnamefont {C.}~\bibnamefont {Anderson}}, \bibinfo
  {author} {\bibfnamefont {J.}~\bibnamefont {Elson}}, \bibinfo {author}
  {\bibfnamefont {J.}~\bibnamefont {Durkee}}, \bibinfo {author} {\bibfnamefont
  {R.}~\bibnamefont {Johns}}, \bibinfo {author} {\bibfnamefont
  {G.}~\bibnamefont {McKinney}}, \bibinfo {author} {\bibfnamefont
  {G.}~\bibnamefont {McMath}}, \bibinfo {author} {\bibfnamefont
  {J.}~\bibnamefont {Hendricks}}, \bibinfo {author} {\bibfnamefont
  {D.}~\bibnamefont {Pelowitz}}, \bibinfo {author} {\bibfnamefont
  {R.}~\bibnamefont {Prael}}, \bibinfo {author} {\bibfnamefont
  {T.}~\bibnamefont {Booth}}, \bibinfo {author} {\bibfnamefont
  {M.}~\bibnamefont {James}}, \bibinfo {author} {\bibfnamefont
  {M.}~\bibnamefont {Fensin}}, \bibinfo {author} {\bibfnamefont
  {T.}~\bibnamefont {Wilcox}},\ and\ \bibinfo {author} {\bibfnamefont
  {B.}~\bibnamefont {Kiedrowski}},\ }\href@noop {} {\emph {\bibinfo {title}
  {{MCNP Users Manual - Code Version 6.2}}}},\ \bibinfo {type} {Tech. Rep.}\
  \bibinfo {number} {LA-UR-17-29981}\ (\bibinfo  {institution} {Los Alamos
  National Laboratory},\ \bibinfo {year} {2017})\BibitemShut {NoStop}%
\bibitem [{\citenamefont {Brown}\ \emph {et~al.}(2018)\citenamefont {Brown},
  \citenamefont {Chadwick}, \citenamefont {Capote}, \citenamefont {Kahler},
  \citenamefont {Trkov}, \citenamefont {Herman}, \citenamefont {Sonzogni},
  \citenamefont {Danon}, \citenamefont {Carlson}, \citenamefont {Dunn},
  \citenamefont {Smith}, \citenamefont {Hale}, \citenamefont {Arbanas},
  \citenamefont {Arcilla}, \citenamefont {Bates}, \citenamefont {Beck},
  \citenamefont {Becker}, \citenamefont {Brown}, \citenamefont {Casperson},
  \citenamefont {Conlin}, \citenamefont {Cullen}, \citenamefont {Descalle},
  \citenamefont {Firestone}, \citenamefont {Gaines}, \citenamefont {Guber},
  \citenamefont {Hawari}, \citenamefont {Holmes}, \citenamefont {Johnson},
  \citenamefont {Kawano}, \citenamefont {Kiedrowski}, \citenamefont {Koning},
  \citenamefont {Kopecky}, \citenamefont {Leal}, \citenamefont {Lestone},
  \citenamefont {Lubitz}, \citenamefont {M\'{a}rquez~Dami\'{a}n}, \citenamefont
  {Mattoon}, \citenamefont {McCutchan}, \citenamefont {Mughabghab},
  \citenamefont {Navratil}, \citenamefont {Neudecker}, \citenamefont {Nobre},
  \citenamefont {Noguere}, \citenamefont {Paris}, \citenamefont {Pigni},
  \citenamefont {Plompen}, \citenamefont {Pritychenko}, \citenamefont
  {Pronyaev}, \citenamefont {Roubtsov}, \citenamefont {Rochman}, \citenamefont
  {Romano}, \citenamefont {Schillebeeckx}, \citenamefont {Simakov},
  \citenamefont {Sin}, \citenamefont {Sirakov}, \citenamefont {Sleaford},
  \citenamefont {Sobes}, \citenamefont {Soukhovitskii}, \citenamefont {Stetcu},
  \citenamefont {Talou}, \citenamefont {Thompson}, \citenamefont {van~der
  Marck}, \citenamefont {Welser-Sherrill}, \citenamefont {Wiarda},
  \citenamefont {White}, \citenamefont {Wormald}, \citenamefont {Wright},
  \citenamefont {Zerkle}, \citenamefont {\v{Z}erovnik},\ and\ \citenamefont
  {Zhu}}]{ENDF8}%
  \BibitemOpen
  \bibfield  {author} {\bibinfo {author} {\bibfnamefont {D.~A.}\ \bibnamefont
  {Brown}}, \bibinfo {author} {\bibfnamefont {M.~B.}\ \bibnamefont {Chadwick}},
  \bibinfo {author} {\bibfnamefont {R.}~\bibnamefont {Capote}}, \bibinfo
  {author} {\bibfnamefont {A.~C.}\ \bibnamefont {Kahler}}, \bibinfo {author}
  {\bibfnamefont {A.}~\bibnamefont {Trkov}}, \bibinfo {author} {\bibfnamefont
  {M.~W.}\ \bibnamefont {Herman}}, \bibinfo {author} {\bibfnamefont {A.~A.}\
  \bibnamefont {Sonzogni}}, \bibinfo {author} {\bibfnamefont {Y.}~\bibnamefont
  {Danon}}, \bibinfo {author} {\bibfnamefont {A.~D.}\ \bibnamefont {Carlson}},
  \bibinfo {author} {\bibfnamefont {M.}~\bibnamefont {Dunn}}, \bibinfo {author}
  {\bibfnamefont {D.~L.}\ \bibnamefont {Smith}}, \bibinfo {author}
  {\bibfnamefont {G.~M.}\ \bibnamefont {Hale}}, \bibinfo {author}
  {\bibfnamefont {G.}~\bibnamefont {Arbanas}}, \bibinfo {author} {\bibfnamefont
  {R.}~\bibnamefont {Arcilla}}, \bibinfo {author} {\bibfnamefont {C.~R.}\
  \bibnamefont {Bates}}, \bibinfo {author} {\bibfnamefont {B.}~\bibnamefont
  {Beck}}, \bibinfo {author} {\bibfnamefont {B.}~\bibnamefont {Becker}},
  \bibinfo {author} {\bibfnamefont {F.}~\bibnamefont {Brown}}, \bibinfo
  {author} {\bibfnamefont {R.~J.}\ \bibnamefont {Casperson}}, \bibinfo {author}
  {\bibfnamefont {J.}~\bibnamefont {Conlin}}, \bibinfo {author} {\bibfnamefont
  {D.~E.}\ \bibnamefont {Cullen}}, \bibinfo {author} {\bibfnamefont {M.~A.}\
  \bibnamefont {Descalle}}, \bibinfo {author} {\bibfnamefont {R.}~\bibnamefont
  {Firestone}}, \bibinfo {author} {\bibfnamefont {T.}~\bibnamefont {Gaines}},
  \bibinfo {author} {\bibfnamefont {K.~H.}\ \bibnamefont {Guber}}, \bibinfo
  {author} {\bibfnamefont {A.~I.}\ \bibnamefont {Hawari}}, \bibinfo {author}
  {\bibfnamefont {J.}~\bibnamefont {Holmes}}, \bibinfo {author} {\bibfnamefont
  {T.~D.}\ \bibnamefont {Johnson}}, \bibinfo {author} {\bibfnamefont
  {T.}~\bibnamefont {Kawano}}, \bibinfo {author} {\bibfnamefont {B.~C.}\
  \bibnamefont {Kiedrowski}}, \bibinfo {author} {\bibfnamefont {A.~J.}\
  \bibnamefont {Koning}}, \bibinfo {author} {\bibfnamefont {S.}~\bibnamefont
  {Kopecky}}, \bibinfo {author} {\bibfnamefont {L.}~\bibnamefont {Leal}},
  \bibinfo {author} {\bibfnamefont {J.~P.}\ \bibnamefont {Lestone}}, \bibinfo
  {author} {\bibfnamefont {C.}~\bibnamefont {Lubitz}}, \bibinfo {author}
  {\bibfnamefont {J.~I.}\ \bibnamefont {M\'{a}rquez~Dami\'{a}n}}, \bibinfo
  {author} {\bibfnamefont {C.~M.}\ \bibnamefont {Mattoon}}, \bibinfo {author}
  {\bibfnamefont {E.~A.}\ \bibnamefont {McCutchan}}, \bibinfo {author}
  {\bibfnamefont {S.}~\bibnamefont {Mughabghab}}, \bibinfo {author}
  {\bibfnamefont {P.}~\bibnamefont {Navratil}}, \bibinfo {author}
  {\bibfnamefont {D.}~\bibnamefont {Neudecker}}, \bibinfo {author}
  {\bibfnamefont {G.~P.~A.}\ \bibnamefont {Nobre}}, \bibinfo {author}
  {\bibfnamefont {G.}~\bibnamefont {Noguere}}, \bibinfo {author} {\bibfnamefont
  {M.}~\bibnamefont {Paris}}, \bibinfo {author} {\bibfnamefont {M.~T.}\
  \bibnamefont {Pigni}}, \bibinfo {author} {\bibfnamefont {A.~J.}\ \bibnamefont
  {Plompen}}, \bibinfo {author} {\bibfnamefont {B.}~\bibnamefont
  {Pritychenko}}, \bibinfo {author} {\bibfnamefont {V.~G.}\ \bibnamefont
  {Pronyaev}}, \bibinfo {author} {\bibfnamefont {D.}~\bibnamefont {Roubtsov}},
  \bibinfo {author} {\bibfnamefont {D.}~\bibnamefont {Rochman}}, \bibinfo
  {author} {\bibfnamefont {P.}~\bibnamefont {Romano}}, \bibinfo {author}
  {\bibfnamefont {P.}~\bibnamefont {Schillebeeckx}}, \bibinfo {author}
  {\bibfnamefont {S.}~\bibnamefont {Simakov}}, \bibinfo {author} {\bibfnamefont
  {M.}~\bibnamefont {Sin}}, \bibinfo {author} {\bibfnamefont {I.}~\bibnamefont
  {Sirakov}}, \bibinfo {author} {\bibfnamefont {B.}~\bibnamefont {Sleaford}},
  \bibinfo {author} {\bibfnamefont {V.}~\bibnamefont {Sobes}}, \bibinfo
  {author} {\bibfnamefont {E.~S.}\ \bibnamefont {Soukhovitskii}}, \bibinfo
  {author} {\bibfnamefont {I.}~\bibnamefont {Stetcu}}, \bibinfo {author}
  {\bibfnamefont {P.}~\bibnamefont {Talou}}, \bibinfo {author} {\bibfnamefont
  {I.}~\bibnamefont {Thompson}}, \bibinfo {author} {\bibfnamefont
  {S.}~\bibnamefont {van~der Marck}}, \bibinfo {author} {\bibfnamefont
  {L.}~\bibnamefont {Welser-Sherrill}}, \bibinfo {author} {\bibfnamefont
  {D.}~\bibnamefont {Wiarda}}, \bibinfo {author} {\bibfnamefont
  {M.}~\bibnamefont {White}}, \bibinfo {author} {\bibfnamefont {J.~L.}\
  \bibnamefont {Wormald}}, \bibinfo {author} {\bibfnamefont {R.~Q.}\
  \bibnamefont {Wright}}, \bibinfo {author} {\bibfnamefont {M.}~\bibnamefont
  {Zerkle}}, \bibinfo {author} {\bibfnamefont {G.}~\bibnamefont
  {\v{Z}erovnik}},\ and\ \bibinfo {author} {\bibfnamefont {Y.}~\bibnamefont
  {Zhu}},\ }\bibfield  {title} {\bibinfo {title} {Endf/b-viii.0: The 8th major
  release of the nuclear reaction data library with cielo-project cross
  sections, new standards and thermal scattering data},\ }\href
  {https://doi.org/10.1016/j.nds.2018.02.001} {\bibfield  {journal} {\bibinfo
  {journal} {Nuclear Data Sheets}\ }\textbf {\bibinfo {volume} {148}},\
  \bibinfo {pages} {1 } (\bibinfo {year} {2018})}\BibitemShut {NoStop}%
\end{thebibliography}%

\end{document}